\documentclass[preprint]{aastex631}

\usepackage{amsmath}
\usepackage{amssymb}
\usepackage{multirow}
\usepackage{graphicx} 
\usepackage{hhline}

\begin{document}

\title{Robust field-level likelihood-free inference with galaxies}

\correspondingauthor{Natalí S. M. de Santi}
\email{natalidesanti@gmail.com}

\author[0000-0002-4728-6881]{Natalí S. M. de Santi}
\affiliation{Center for Computational Astrophysics, Flatiron Institute, 162 5th Avenue, 
New York, NY, 10010, USA}
\affiliation{Instituto de Física, Universidade de São Paulo, R. do Matão 1371, 05508-900, 
São Paulo, Brasil}

\author[0000-0002-0152-6747]{Helen Shao}
\affiliation{Department of Astrophysical Sciences, Princeton University, 4 Ivy Lane, Princeton, 
NJ 08544 USA}

\author[0000-0002-4816-0455]{Francisco Villaescusa-Navarro}
\affiliation{Center for Computational Astrophysics, Flatiron Institute, 162 5th Avenue, 
New York, NY, 10010, USA}
\affiliation{Department of Astrophysical Sciences, Princeton University, 4 Ivy Lane, Princeton, 
NJ 08544 USA}

\author[0000-0001-8295-7022]{L. Raul Abramo}
\affiliation{Instituto de Física, Universidade de São Paulo, R. do Matão 1371, 05508-900, 
São Paulo, Brasil}

\author[0000-0001-7689-0933]{Romain Teyssier}
\affiliation{Department of Astrophysical Sciences, Princeton University, 4 Ivy Lane, Princeton, NJ 08544 USA}

\author[0000-0002-0936-4279]{Pablo Villanueva-Domingo}
\affiliation{Computer Vision Center - Universitat Aut\`onoma de Barcelona (UAB), 08193 
Bellaterra, Barcelona, Spain}

\author[0000-0001-7899-7195]{Yueying Ni}
\affiliation{Harvard-Smithsonian Center for Astrophysics, 60 Garden Street, Cambridge, MA 
02138, US}
\affiliation{McWilliams Center for Cosmology, Department of Physics, Carnegie Mellon 
University, Pittsburgh, PA 15213, US}

\author[0000-0001-5769-4945]{Daniel Anglés-Alcázar}
\affiliation{Department of Physics, University of Connecticut, 196 Auditorium Road, U-3046, 
Storrs, CT, 06269, USA}
\affiliation{Center for Computational Astrophysics, Flatiron Institute, 162 5th Avenue, 
New York, NY, 10010, USA}

\author[0000-0002-3185-1540]{Shy Genel}
\affiliation{Center for Computational Astrophysics, Flatiron Institute, 162 5th Avenue, 
New York, NY, 10010, USA}
\affiliation{Columbia Astrophysics Laboratory, Columbia University, 550 West 120th Street, 
New York, NY 10027, USA}

\author[0000-0002-1329-9246]{Elena Hernandez-Martinez}
\affiliation{Universit\"ats-Sternwarte, Fakult\"at f\"ur Physik, Ludwig-Maximilians-Universit\"at M\"unchen, Scheinerstr. 1, 81679 M\"unchen, Germany}

\author[0000-0001-8867-5026]{Ulrich P. Steinwandel}
\affiliation{Center for Computational Astrophysics, Flatiron Institute, 162 5th Avenue, 
New York, NY, 10010, USA}

\author[0000-0001-7964-5933]{Christopher C. Lovell}
\affiliation{Institute of Cosmology and Gravitation, University of Portsmouth, Burnaby Road, 
Portsmouth, PO1 3FX, UK}
\affiliation{Centre for Astrophysics Research, School of Physics, Engineering \& Computer 
Science, University of Hertfordshire, Hatfield AL10 9AB, UK}

\author{Klaus Dolag}
\affiliation{Universit\"ats-Sternwarte, Fakult\"at f\"ur Physik, Ludwig-Maximilians-Universit\"at M\"unchen, Scheinerstr. 1, 81679 M\"unchen, Germany}
\affiliation{Max-Planck-Institut f\"ur Astrophysik, Karl-Schwarzschild-Stra{\ss}e 1, 85741 Garching, Germany}

\author[0000-0002-6292-3228]{Tiago Castro}
\affiliation{INAF-Osservatorio Astronomico di Trieste, Via G. B. Tiepolo 11, I-34143 Trieste, 
Italy}
\affiliation{INFN, Sezione di Trieste, Via Valerio 2, I-34127 Trieste TS, Italy}
\affiliation{IFPU, Institute for Fundamental Physics of the Universe, via Beirut 2, 34151 
Trieste, Italy}

\author{Mark Vogelsberger}
\affiliation{Kavli Institute for Astrophysics and Space Research, Department of Physics, MIT, 
Cambridge, MA 02139, USA}
\affiliation{The NSF AI Institute for Artificial Intelligence and Fundamental Interactions, 
Massachusetts Institute of Technology, Cambridge MA 02139, USA}

\begin{abstract}

We train graph neural networks to perform field-level likelihood-free inference using galaxy catalogs from
state-of-the-art hydrodynamic simulations of the CAMELS project. 
Our models are rotational, translational, and permutation invariant and do not impose any cut on scale. 
From galaxy catalogs that only contain $3$D positions and radial velocities of $\sim 1, 000$  galaxies in
tiny $(25~h^{-1}{\rm Mpc})^3$ volumes our models can infer the value of $\Omega_{\rm m}$ with approximately 
$12$\% precision.
More importantly, by testing the models on galaxy catalogs from thousands of hydrodynamic simulations, each
having a different efficiency of supernova and AGN feedback, run with five different codes and subgrid models
-- IllustrisTNG, SIMBA, Astrid, Magneticum, SWIFT-EAGLE --, we find that our models are robust to changes in
astrophysics, subgrid physics, and subhalo/galaxy finder.
Furthermore, we test our models on 1,024 simulations that cover a vast region in parameter space 
-- variations in 5 cosmological and 23 astrophysical parameters -- finding that the model extrapolates really 
well. 
Our results indicate that the key to building a robust model is the use of both galaxy positions 
and velocities, suggesting that the network have likely learned an underlying physical relation that does not 
depend on galaxy formation and is valid on scales larger than $\sim10~h^{-1}{\rm kpc}$.

\end{abstract}


\section{Introduction} 
\label{sec:intro}

The standard model of Cosmology describes a Universe filled with dark matter (DM), baryonic 
matter and some form of dark energy (DE).
Despite many observational constraints, such as the temperature and polarization fluctuations 
in the cosmic microwave background \citep{Bennett2013, planck2018}, many mysteries still 
remain, in particular, the fundamental natures of DE and DM.
In order to solve the remaining puzzles and consolidate this physical description, cosmologists
aim at constraining, with the highest precision and accuracy possible, the parameters of the
model.

Since the distribution of matter and galaxies in the Universe depends on the
cosmological parameters, the clustering of these objects can be used to infer the
values of those parameters. 
In order to collect as much data as diversely as possible large international 
efforts are currently underway to survey the cosmos at different wavelengths:  
DESI \citep{DESI}, 
Euclid \citep{Laureijs2011, Amendola2012, Euclid2016, Euclid2022-Tiago_Castro}, 
Prime Focus Spectrograph (PFS) \citep{Ellis2012}, J-PAS \citep{Benitez2014},
Square Kilometer Array (SKA) \citep{SKA1999}, 
Roman \citep{Roman2015}, JWST \citep{JWST}, and others. 
The data from these missions will encompass larger volumes at different redshifts, 
using a variety of different types of galaxies, observed at many wavelengths.
Extracting the maximum amount of relevant information from these data sets is of key 
importance in order to improve our understanding of fundamental physics.

To achieve that goal theoretical predictions and methods to extract that information are needed. 
On the one hand, we have traditional methods for extracting information from cosmological observations.
In the case of Bayesian inference schemes for  cosmological parameters, nearly all
analyses make use of summary statistics, like the power spectrum.
However, this approach is sub-optimal as we do not know what summary statistics contain all 
(or the majority) of the cosmological information 
\citep{Hahn2020, Uhlemann2020, Gualdi2021, Banerjee2021}. 
Furthermore, usual methods normally require expensive simulations to either 
estimate covariance matrices or to forward-model the observations 
\citep{Efron1980, Taylor2013, Alan2018, CARPool2022, deSanti2022JCAP}.

On the other hand, machine learning (ML) techniques have been shown to outperform 
traditional methods in a large variety of contexts and areas, including cosmology and 
astrophysics. 
In fact, the power of these new methods resides precisely in their ability to deal with 
large and complex data sets, providing  nonlinear relations in high-dimensional 
feature spaces that allow us to solve regression and classification tasks 
\citep{ivezic2014statistics}. 
Using different summary statistics as input data, \cite{lucia2022} are 
able to derive cosmological parameters without the need for additional input from 
theoretical models, thus providing a powerful generalization of the usual Monte 
Carlo-based methods.
In particular, likelihood-free inference methods work by taking data directly from the 
simulations (without the need for summary statistics and, thus, model comparison), 
and many papers have shown competitive results compared with the usual statistical 
inference methods 
\citep{Ravanbakhsh2017, Ntampaka2020, Mangena2020, Hassan2020, Paco2021, Cole2021, pablo-galaxies-2022, 
lucas2022, helen-halos-2022}. 
At a level closer to the observations and simulations, many papers exploring the halo--galaxy connection are
able to make predictions that are comparable to the output of 
numerical/analytical methods 
\citep{Jo2019, Yip2019, Zhang2019, Kamdar2016, Wadekar2020, Kasmanoff2020, Moster2021, McGibbon2021, Shao2021, vonMarttens2021, pablo-via_lactea-2021,Delgado2021, deSanti2022MNRAS, Christian2022, pablo-halo_mass-2022, Lovell2022, Rodrigues2023}. 
Furthermore, a clear advantage of ML models is that, once they are trained, they typically 
make predictions much faster than traditional methods \citep{Christian2022} and a disadvantage 
arises when these models fail to extrapolate their predictions across different
data sets, from the ones with which they were trained with 
\citep{Paco2021, pablo-galaxies-2022, one_gal-2022}.

Machine learning algorithms can also work with sparse and irregular data; e.g. through 
graph neural networks (GNNs) \citep{Gilmer2017, Battaglia2018, Bronstein2021}. 
GNNs exhibit multiple advantages over convolutional neural networks (CNNs). 
For instance, in the context of cosmology, they do not impose any cut on the 
considered physical scales, and different physical symmetries (e.g. translational and 
rotational invariance) can be easily implemented in the models 
\citep[see][]{pablo-galaxies-2022}. 
GNNs have been used for a variety of tasks, such as parameter inference 
\citep{pablo-galaxies-2022, helen-halos-2022, lucas2022, Anagnostidis_2022}, inferring 
halo masses \citep{pablo-via_lactea-2021}, speeding up semi-analytic models 
\citep{Christian2022}, and rediscovering Newton's law \citep{Cramer2020}. 

In particular, \cite{pablo-galaxies-2022} showed that GNNs were able to infer 
$\Omega_{\rm m}$ with $\sim 10 \%$ accuracy just based on galaxy properties (e.g., 
positions, stellar mass, radius, and metallicity), without making use of any summary 
statistics and performing likelihood-free inference.
However, their model was not robust.
The lack of robustness in this field-level inference task with galaxies through GNNs 
can be attributed to many reasons: from intrinsic differences in the subgrid models 
of the different simulations to the models learning unique, numerical, artifacts. 
Developing robust models, that extrapolate properly even with real data, is one most important tasks
needed to replace standard data analysis techniques (e.g. perturbation theory).
\cite{helen-halos-2022} showed instead that the positions and velocities of DM halos were 
robust to numerics in N-body codes as well as to variations in astrophysical parameters when 
inferring $\Omega_{\rm m}$ using a field-level approach. 
Even so, this work still deals with non-observables, such as DM halos, and some of their properties.
In our companion paper, \cite{helen-prep} we are providing analytical equations to predict 
$\Omega_{\rm m}$ from the positions and velocity modulus fields of DM halos. 
This model is robust across different DM N-body simulations and, by changing the normalization
of the input velocity modulus for each hydrodynamic simulation, it is able to perform 
predictions for galaxy catalogs too. These equations bring a big step towards a physical interpretation 
of the model.

In the present work, we extend all these previous efforts using GNNs to show that we can 
build models that perform field-level likelihood-free inference using galaxy catalogs 
that are robust to changes in numerics, astrophysics, subgrid physics, and the 
method to identify galaxies. 
We train GNNs using thousands of galaxy catalogs from state-of-the-art hydrodynamic
simulations of the CAMELS project \citep{Paco2021-projCAMELS}.
We also investigate which galaxy properties are robust and how they contribute to the 
network predictions, showing that we only need the phase-space information of the galaxies to achieve
the best results.

The manuscript is organized as follows: in Section \ref{sec:data} we present the data 
set, describing the different simulations used and their different setups; 
in Section \ref{sec:methodology} we describe the methodology, 
where we explain the data pre-processing, the translation of the galaxy catalogs 
into graphs, and the general architecture employed; in Section \ref{sec:results}, we 
present the results related to the best model, our efforts to improve it, and investigate
which is the most important source of information for the GNNs to extract their 
inferences; and, in Section \ref{sec:conclusions}, we present the discussion and 
conclusions, analyzing the differences among the different simulations, and provide 
ideas for future work.

\begin{table*}
 \caption{\label{tab:summary} Characteristics of the hydrodynamical simulations used in this work.}
 \begin{center}
 \resizebox{\textwidth}{!}{%
  \begin{tabular}{ccccc}
   \hline\hline
   \multirow{3}{*}{\textbf{Model}} & \multirow{3}{*}{\textbf{Usage}} & 
   {\bf Number of} & \textbf{Mean number} & \multirow{3}{*}{{\bf Reference}} \\
   & & {\bf simulations} & \textbf{of galaxies} & \\
   & & {\bf used} & \textbf{per catalog} & \\
   \hline\hline
    Astrid & Train, validate \& test & 1000(LH) + 27(CV) & 1114 & 
    \cite{Astrid2022} \\
    SIMBA        & Train, validate \& test & 1000(LH) + 27(CV) & 1093 & 
    \cite{SIMBA2019}                 \\
    IllustrisTNG & Train, validate \& test & 1000(LH) + 27(CV) + 1024(SB) & 737  &
    \cite{Pillepich2018}             \\
    IllustrisTNG300 & Test & 1(LH) & 799  &
    \cite{Nelson2019}             \\
    Magneticum   & Test          & 50(LH) + 27(CV)   & 3655 & \cite{MAGNETICUM2014}            \\
    SWIFT-EAGLE        & Test          & 64(LH)             & 1255 & \cite{EAGLE2015}            \\
   \cline{1-5}
   \end{tabular}}
  \end{center}
\end{table*}

\section{Data} 
\label{sec:data}

In this section, we describe the data we use to train, validate, and test our models.
We emphasize that all the galaxy properties considered in this work are direct from the simulations.
In this way, we are not performing any changes in order to consider realistic effects, such as taking into
account errors in the peculiar velocities. These considerations will be addressed in future work.

\subsection{Simulations}

The galaxy catalogs we use to train, validate, and test our models come from thousands 
of hydrodynamic simulations of the Cosmology and Astrophysics with MachinE Learning 
Simulations -- CAMELS project \citep{Paco2021-projCAMELS, Villaescusa-Navarro2022-relCAMELS}. 
The hydrodynamic simulations have been run with different codes that solve the 
hydrodynamic equations differently and implement different subgrid models: 
IllustrisTNG \citep{Weinberger2016,Pillepich2018}, SIMBA \citep{SIMBA2019}, Astrid \citep{Astrid2022},
Magneticum \citep{MAGNETICUM2014}, and SWIFT-EAGLE \citep{EAGLE2015, Schaller2016}. 
All the simulations follow the evolution of $256^3$ DM particles and are initialized with $256^3$ 
fluid elements from $z = 127$ down to $z = 0$ in periodic boxes of $25~h^{-1}{\rm Mpc}$ on a side. 
The catalogs used in this work correspond to $z = 0$.  
The fiducial values of the cosmological parameters are: $\Omega_{\rm m} = 0.3$, 
$\Omega_{\rm b} = 0.049$, $h = 0.6711$, $n_s = 0.9624$, $\sigma_8 = 0.8$, $w = - 1$, 
$M_\nu = 0$ eV. 

The CAMELS simulations can be classified into different sets and suites depending on how 
their parameters are arranged and which code was used to run them. We start by classifying 
the catalogs into different sets:
\begin{itemize}
 \item {\bf Latin Hypercube (LH).} The simulations in this category have their 
 cosmological and astrophysical parameter variations arranged in a Latin hypercube that spans: 
 $\Omega_{\rm m} \in [0.1, 0.5]$ and $\sigma_8 \in [0.6, 1.0]$, 
 $A_{\rm SN1} \in [0.25, 4.0]$, $A_{\rm SN2} \in [0.5, 2.0]$, 
 $A_{\rm AGN1} \in [0.25, 4.0]$, and $A_{\rm AGN2} \in [0.5, 2.0]$. $A_{\rm SN}$
 and $A_{\rm AGN}$ are astrophysical parameters that control the efficiency of 
 supernova (SN) and active galactic nuclei (AGN) feedback (see \cite{Paco2021-projCAMELS, 
 Yueying-prep} for a detailed description of the meaning of the astrophysical simulations in 
 every simulation suite).
 Each of the simulations in the Latin hypercube has been run with a different initial random 
 seed for the generation of the initial conditions. We used these simulations for training, validating, and testing.
 
 \item {\bf Cosmic Variance (CV).} These simulations have been run with the fiducial 
 value of the cosmological and astrophysical parameters. The initial conditions for each simulation in this set 
 have been generated with a different initial random seed. These simulations are only used 
 for testing the models.

 \item {\bf Sobol Sequence (SB).} The simulations in this set have their cosmological 
 and astrophysical parameters arranged in a Sobol sequence \citep{SOBOL1967}. 
 A total of $28$ parameters are varied: $5$ cosmological ($\Omega_{\rm m}$, $\Omega_{\rm b}$, 
 $h$, $n_s$, $\sigma_8$) and $23$ astrophysical. 
 The astrophysical parameters varied include the usual ones
 ($A_{\rm SN1}, A_{\rm SN2}, A_{\rm AGN1}, A_{\rm AGN2}$) 
 and incorporate many others such as star formation, galactic winds, black hole (BH) growth 
 and quasar parameters. 
 All of them vary in ranges around the fiducial values used in the IllustrisTNG set. 
 Their range of variation is large enough to enable a broad sampling of the considered 
 parameter \citep{Yueying-prep}. We note that this set covers the largest region in parameter 
 space within CAMELS although at a much lower density given the high dimensionality of the 
 considered space. 
 We use these simulations only for testing and to investigate how well our models generalize.
 
\end{itemize}

The CAMELS simulations can also be classified into different model suites according to the 
code used to run them:

\begin{itemize}
 \item {\bf IllustrisTNG.} These simulations were run using \textsc{Arepo} 
 \citep{springel2010, Weinberger2020} applying the same subgrid physics as the IllustrisTNG 
 simulations \citep{Weinberger2016, Pillepich2018}. This suite 
 contains $1000$ LH, $27$ CV, and $1024$ SB simulations. 
 
 \item {\bf SIMBA.} These simulations were run with the \textsc{Gizmo} code \citep{Hopkins2015}
 and employ the same subgrid physics as the SIMBA simulation \citep{SIMBA2019}. This suite 
 contains $1000$ LH and $27$ CV simulations.
  
 \item {\bf Astrid.} These simulations were run using MP-Gadget \citep{MPGadget} applying some 
 modifications to the subgrid model employed in the Astrid simulation 
 \citep{Astrid-Y-2022, Astrid2022, Yueying-prep}.  This suite contains $1000$ LH and $27$ CV 
 simulations.
 
 \item {\bf Magneticum.} These simulations were run with the parallel cosmological Tree-PM code 
 P-Gadget3 \citep{GADGET2}. The code uses an entropy-conserving formulation of Smoothed 
 Particle Hydrodynamics (SPH) \citep{Springel2002}, with SPH modifications according to 
 \cite{Dolag2004ApJ, Dolag2005MNRAS, Dolag2006MNRAS}. It includes also prescriptions for 
 multiphase interstellar medium based on the model by \cite{Springel2003MNRAS} as well as
 \cite{Tornatore2007} for the metal enrichment prescription. The model follows the growth and 
 evolution of BHs and their associated AGN feedback based on the model presented by
 \citet{Springel2005MNRAS} and \citet{DiMatteo2005}, but includes modifications based on 
 \citet{Fabjan2011MNRAS}, \citet{Hirschmann2014MNRAS} and \citet{Steinborn2016MNRAS}. The set 
 contains $50$ LH and $27$ CV simulations. The following subgrid parameters were varied in 
 order to control the stellar and AGN feedback (with parameter ranges given in square brackets) 
 on the Latin-hypercube:

   \begin{itemize}
    \item $ A_{SN1}$, energy per unit of SFR [0.25, 4.0] $\times 10^{51}$.
    \item $ A_{SN2}$, wind speed [250,1000].
    \item $ A_{AGN1}$, coupling efficiency of the BH feedback [0.25, 4.0].
    \item $ A_{AGN2}$, boost of the AGN mode feedback [0.5, 2.0].
  \end{itemize}

\item {\bf SWIFT-EAGLE.} These simulations have been run with the \textsc{Swift} code 
\citep{Schaller2016, Schaller2018} using a new subgrid physics model based 
on the original Gadget-EAGLE simulations \citep{EAGLE2015, Crain2015}, with some parameter 
changes \citep{EAGLE-Borrow2022}. 
The full model will be described in \cite{Borrow-prep}. This suite contains $64$ LH simulations 
varying the following subgrid parameters controlling the stellar and AGN feedback (with 
parameter ranges given in square brackets) on the Latin hypercube:
  \begin{itemize}
    \item $f_{\mathrm{E, min}}$, the minimal stellar feedback fraction, [0.18, 0.6].
    \item $f_{\mathrm{E, max}}$, the maximal stellar feedback fraction, [5, 10].
    \item $N_{\mathrm{H, 0}}$, pivot point in density that the feedback energy fraction plane
    rotates around, [$10^{-0.6}$, $10^{-0.15}$].
    \item $\sigma_{\mathrm{n}}$ and $\sigma{\mathrm{Z}}$, energy fraction sigmoid width, 
    controlling the density and metallicity dependence, [0.1, 0.65].
    \item $\varepsilon_{\mathrm{f}}$, coupling coefficient of radiative efficiency of AGN 
    feedback, [$10^{-2}$, $10^{-1}$].
    \item $\Delta T_{\mathrm{AGN}}$, AGN heating temperature, [$10^{8.3}$, $10^{9.0}$].
    \item $\alpha$, BH accretion suppression/enhancement factor, [0.2, 1.1].
  \end{itemize}
\end{itemize}

Finally, to quantify the robustness of our model to super-sample covariance effects, we made 
use of the IllustrisTNG300-1 simulation \citep{Nelson2019}, which covers a larger volume of 
($205~h^{-1}{\rm Mpc}$)$^3$ with slightly higher resolution than our fiducial CAMELS 
simulations and has a slightly different cosmology:  $\Omega_{\rm m} = 0.3089$,
$\Omega_{\rm b} = 0.0486$, $\Omega_\Lambda = 0.6911$, $h = 0.6774$, $\sigma_8 = 0.8159$, and
$n_s = 0.9667$. This simulation was run with \textsc{Arepo} and made use of the IllustrisTNG
subgrid physics model \citep{Weinberger2017, Pillepich2018, Pillepich2018MNRAS, Marinacci2018, Naiman2018, Nelson2018, Springel2018}. 

We emphasize that although the name of the parameters $A_{\rm SN1}, A_{\rm SN2}, A_{\rm AGN1}, 
A_{\rm AGN2}$ is common among different simulations, their actual implementation and effect on 
galaxy properties and clustering can be very distinct. Therefore, it is important to keep in 
mind that those parameters are not meant to share physical effects, only their names.

\subsection{Galaxy catalogs}

Halos and subhalos are identified in the simulations for every snapshot using two different 
halo and subhalo finders: \textsc{SubFind} \citep{Springel2001, Dolag2009} and 
\textsc{VELOCIraptor} \citep{velociraptop1, velociraptor2}. 
All galaxy catalogs are from \textsc{SubFind} with the 
exception of SWIFT-EAGLE, which only contains \textsc{VELOCIraptor} catalogs. The reason for 
using two different codes is to check the robustness of our results to
the subhalo finding procedure, which can cause some differences in the number of galaxies as 
shown in \cite{fight-halo_finders-2022}.

Galaxies are defined in all cases as subhalos that contain at least one star particle. 
In this work, we only consider galaxies with stellar masses above $1.3\times10^8~M_\odot/h$.
A galaxy catalog is constructed by taking all galaxies whose stellar mass is higher than a 
given threshold. For every simulation, we produce several galaxy catalogs by varying the 
stellar mass threshold.

A summary of the simulation characteristics can be found in Table \ref{tab:summary}, where 
we present their usage, the number of catalogs, the mean number of galaxies per catalog
and the reference for each of the original galaxy formation models.

\begin{figure*}
 \centering
 \includegraphics[scale=0.41]{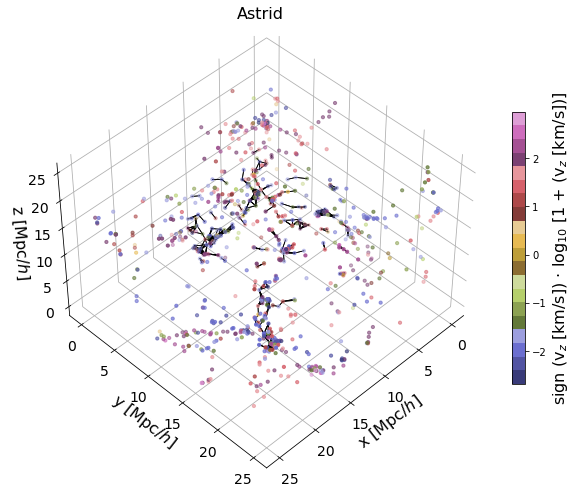}
 \includegraphics[scale=0.41]{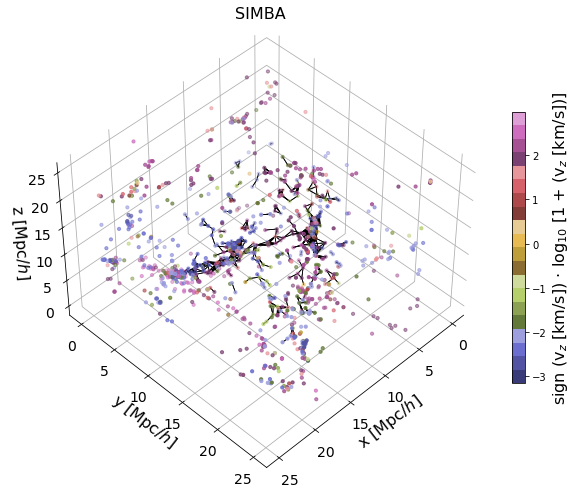}
 \includegraphics[scale=0.41]{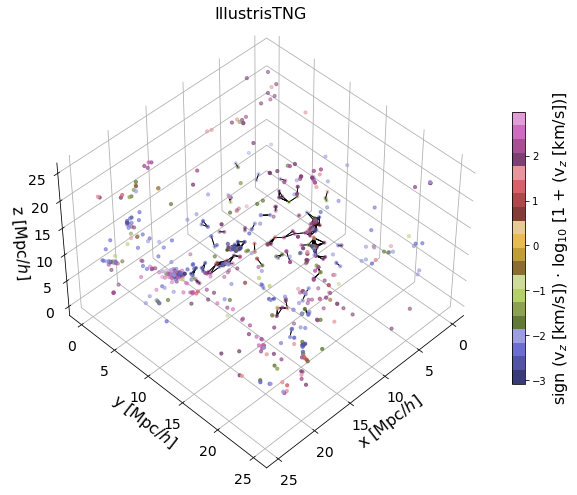}
 \includegraphics[scale=0.41]{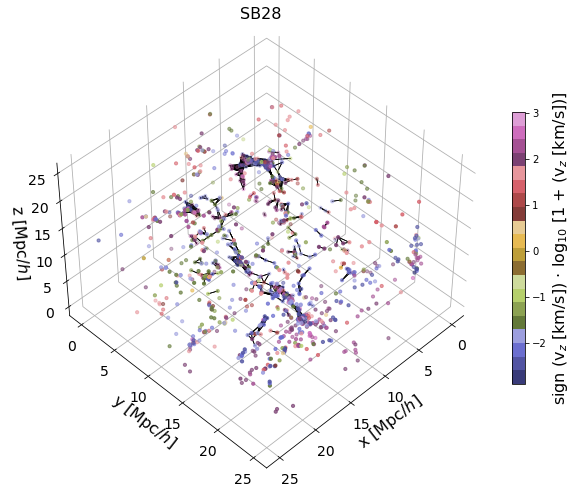}
 \includegraphics[scale=0.41]{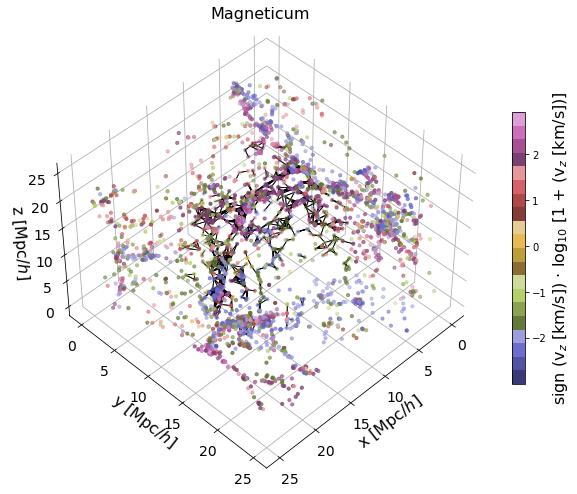}
 \includegraphics[scale=0.41]{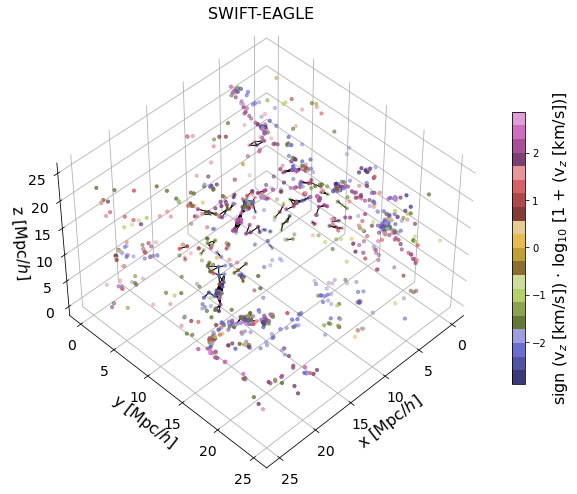}
 \caption{Examples of graphs constructed from galaxy catalogs from different CAMELS 
 simulations: Astrid, SIMBA, IllustrisTNG, Magneticum, SB28, and SWIFT-EAGLE. The nodes 
 represent the galaxies and their colors correspond to the normalization (Equation 
 \ref{eq:norm_vz}) of the $z$ component of their peculiar velocity. Galaxies are connected by 
 edges (shown as black lines) if their distance is smaller than 
 $r_{\mathrm{link}} \sim 1.25~h^{-1}{\rm Mpc}$. We stress that we are no
 galaxies which are linked due to periodic boundary conditions in these plots.}
 \label{fig:graphs}
\end{figure*}

\section{Methodology} 
\label{sec:methodology}

In this section, we describe: the method we use to construct graphs from galaxy catalogs 
(Section \ref{sec:the_graph}); the architecture of our GNN 
(Section \ref{sec:architecture}); the method to carry out likelihood-free inference
(Section \ref{sec:free_likelihood_inference}); the training procedure and optimization 
choices (Section \ref{sec:train_optm}); and the evaluation of the methodology, where we 
present the scores for the metrics we analyzed (Section \ref{sec:scores}).

\subsection{Galaxy graphs: construction}
\label{sec:the_graph}

The input for our GNNs are graphs: mathematical structures characterized by nodes, edges, and
global properties. Every element of the graph can be described by a set of properties: 
$\mathbf{n}_i$ represents the properties of node $i$, $\mathbf{e}_{i j}$ represents the 
features of the edge between node $i$ and $j$, and $\mathbf{g}$ contains global properties 
of the graph \citep{Gilmer2017, Zhou2018, Battaglia2018}. We construct graphs from catalogs 
that contain the galaxy positions and their peculiar velocities (only the $z$ component); in 
some models, we also include the stellar mass of the galaxies.

In this work, we follow the method presented in \cite{pablo-galaxies-2022}
(and used in \cite{helen-halos-2022} and \cite{lucas2022} for halos) 
where galaxies represent the graph nodes and two galaxies are connected by an edge if their 
distance is smaller than a given linking radius $r_{link}$. 
Additionally, we use as a global property of the graph the logarithm of the number of galaxies 
in the graph: $\log_{10} ( N_g )$\footnote{We have checked that including the number of 
galaxies as global feature yields slightly better results. For that reason, we keep that 
property. \label{g_number}}. 

We investigate the contribution of the $z$ component of the galaxy's peculiar velocities $v_z$ 
and the stellar mass $M_{\star}$ as node attributes. We transform these features according to:
\begin{align}
 v_z & \rightarrow \mathrm{sign} (v_z) \cdot \log_{10} \left[ 1 + \mathrm{abs} (v_z) \right] , \label{eq:norm_vz}\\
 M_{\star} & \rightarrow \log_{10} (1 + M_{\star}) ~.
\end{align}
We chose to work with only one component for the galaxy velocity. This is because we 
want to be as close as possible to observational data, where we have access only to the radial 
peculiar velocity, i.e., the velocity measured along the line of sight.

The edge features contain information about the spatial distribution of galaxies (their 
positions), and those properties are designed to make the graph invariant under rotations and 
translations. We follow \citet{pablo-galaxies-2022} and set the edge features as:
\begin{equation}
 \mathbf{e}_{i j} = \left[ \frac{|\mathbf{d}_{i j}|}{r_{link}}, \alpha_{i j}, \beta_{i j} \right] \, ,
\end{equation}
where:
\begin{align}
 \mathbf{d}_{i j} & = \left[ \mathbf{r}_{i} - \mathbf{r}_{j} \right]\\
 \pmb{\delta}_{i} & = \mathbf{r}_{i} - \mathbf{c}\\
 \alpha_{i j} & = \frac{ \pmb{\delta}_{i}}{ |\pmb{\delta}_{i}| } \cdot \frac{ \pmb{\delta}_{j}}{ |\pmb{\delta}_{j}| }\\   
 \beta_{i j} & = \frac{ {\pmb \delta}_{i}}{ |{\pmb\delta}_{i}| } \cdot \frac{ \mathbf{d}_{i j}}{ |\mathbf{d}_{i j}| } \, ,
\end{align}
with $\mathbf{r}_{i}$ representing the position of a galaxy $i$ and 
$\mathbf{c}=\sum_i^N \mathbf{r}_i/N$ being
the centroid.
Here, the {\em distance} $\mathbf{d}_{i j}$ is the difference of two galaxy ($i$ and $j$)
positions, the {\em difference vector} $\pmb{\delta}_{i}$ denotes the position of a galaxy 
$i$ with respect to the centroid, $\alpha_{i j}$ is the (cosine of) the angle between the
difference vectors of two galaxies,  while $\beta_{i j}$ represents the angle between the
difference vector of a galaxy $i$ and its distance to another galaxy $j$. 
We account for periodic boundary conditions when computing both distances and angles. 
Moreover, we consider reverse edges -- a copy of the graphs, with the same
nodes and edges but with all of the edges reversed while compared to the
orientation of the corresponding edges in the original graph; we do not consider 
self-loops (an edge that connects a node to itself). Note that, by construction, the model
is rotational and translation invariant, as those operations will not change the  
edge features of the graph. In other words, they will remain the same while 
performing the usual rotation and translational matrix transformations to the galaxy 
positions \citep{pablo-galaxies-2022}.

In Figure \ref{fig:graphs} we show graphs constructed from galaxy catalogs of the
different simulations: Astrid, SIMBA, IllustrisTNG, SB28, Magneticum, and SWIFT-EAGLE. 
All these catalogs contain galaxies with minimum stellar mass: 
$M_{\star} = 1.95 \times 10^8~M_\odot/h$.
In all the graphs galaxies are colored according to their $v_z$  
(transformed according to Equation \ref{eq:norm_vz}), and two galaxies are connected by a black 
line if their distance is within $r_{\mathrm{link}} \simeq 1.25~h^{-1}{\rm Mpc}$ (this value 
was found with \textsc{Optuna}, as it will be described in Section \ref{sec:train_optm}). 
Notice that we are not 
connecting galaxies which are linked due to the periodic boundary conditions in this 
representation, i.e., a galaxy near the border of the box is not showing to be connected
to some other galaxy in the other box extreme, even when they are linked due to these
conditions.
This simple visual comparison shows that the spatial distribution of galaxies and their
peculiar velocities are similar among all simulations. We note that the graph constructed from 
the Magneticum simulation exhibits a significantly larger number of galaxies than the 
others; this happens due to the employed AGN model used in Magneticum.

Every graph is characterized by a set of labels that we aim at inferring 
(e.g. $\Omega_{\rm m}$). We normalize these labels as  $\theta_i$, using
\begin{equation}
 \theta_i \rightarrow \frac{ \left( \theta_i - \theta_{\rm min} \right) }{ \left( \theta_{\rm max} - \theta_{\rm min} \right) } ,   
\end{equation}
where $\theta_{\rm min}$ and $\theta_{\rm max}$ represent the minimum and the maximum values 
of the corresponding parameter.

More details about the construction of the graphs, as well as an analysis of the different
graphs (for the different simulations) are presented in Appendix \ref{sec:graph_details}.

\subsection{GNN architecture}
\label{sec:architecture}

The architecture we employ in this work follows the one presented in \textsc{CosmoGraphNet}\footnote{Available on \textsc{Github} repository \href{https://github.com/PabloVD/CosmoGraphNet}{https://github.com/PabloVD/CosmoGraphNet}, DOI: \href{https://doi.org/10.5281/zenodo.6485804}{10.5281/zenodo.6485804}.}
\citep{pablo-galaxies-2022}. 
Basically, the GNN is trained to infer the value of some cosmological parameter 
($\Omega_{\rm m}$) from an input graph. Because GNNs are designed to deal with irregular 
and sparse data, the main idea behind their work was to perform a transformation of their
components information (nodes $\mathbf{n}_i$, edges $\mathbf{e}_{i j}$, and global 
$\mathbf{g}$ attributes are updated), while the graph structure is preserved. 
In the end, the information is compressed, being converted by a usual multi-layer 
perceptron (MLP), to deliver the final property of the graph. By construction, GNNs preserve the
graph symmetries (permutational invariance in the nodes, edge, and global attributes 
\citep{Gilmer2017, Battaglia2018, Bronstein2021}). Besides, as done in 
\cite{pablo-galaxies-2022}, the edge attributes consider translational and rotational symmetries
(and here account for periodic boundary conditions too).

We have used a {\em message passing scheme} where each message passing layer updates the 
node and edge features\footnote{Note that we do not employ a model to update the global 
attribute. 
This was used just in order to update the node information (see Equation 
\ref{eq:node_model}).}, 
taking as input the graph and delivering as output its updated version. The node and edge
features at layer $\ell + 1$ are found from the node and edge features at layer $\ell$ as:

\begin{itemize}

 \item {\bf Edge model:}
  \begin{equation}
   \mathbf{e}_{i j}^{(\ell + 1)} = \mathcal{E}^{(\ell + 1)} \left( \left[ \mathbf{n}_{i}^{(\ell)}, 
   \mathbf{n}_{j}^{(\ell)}, \mathbf{e}_{i j}^{(\ell)} \right]  \right) ,
   \label{eq:edge_model}
  \end{equation}
  where $\mathcal{E}^{(\ell + 1)}$ represents a MLP;
 
 \item {\bf Node model:}
  \begin{equation}
   \mathbf{n}_{i}^{(\ell + 1)} = \mathcal{N}^{(\ell + 1)} \left( \left[ \mathbf{n}_{i}^{(\ell)}, 
   \bigoplus_{j \in \mathfrak{N}_i} \mathbf{e}_{i j}^{(\ell + 1)}, \mathbf{g} \right]  \right) ,
   \label{eq:node_model}
  \end{equation}
  where $\mathfrak{N}_i$ represents all neighbors of node $i$, $\mathcal{N}^{(\ell + 1)}$ is 
  a MLP, and $\oplus$ is a multi-pooling operation responsible to concatenate several 
  permutation invariant operations:
  \begin{equation}
   \bigoplus_{j \in \mathfrak{N}_i} \mathbf{e}_{i j}^{(\ell + 1)} = \left[ 
   \max_{j \in \mathfrak{N}_i} \mathbf{e}_{i j}^{(\ell + 1)},
   \sum_{j \in \mathfrak{N}_i} \mathbf{e}_{i j}^{(\ell + 1)},
   \frac{ \sum_{j \in \mathfrak{N}_i} \mathbf{e}_{i j}^{(\ell + 1)} }{ \sum_{j \in \mathfrak{N}_i} }
   \right] .  \label{eq:multi-pooling}
  \end{equation}
\end{itemize}
The use of the multi-pooling operation in the equation above was made because
it has been argued that several aggregators can enhance the expressiveness of GNNs \citep{Corso2020}.
Additionally, the number of layers to perform this update is a hyperparameter to be chosen 
in the optimization scheme. 
We also we made use of residual layers in the intermediate layers. The use of residuals means
adding the input of the layer to its respective output, i.e., adding node/edge attributes to
node/edge models. A discussion about this use can be found in \cite{Li2017} and \cite{pablo-galaxies-2022}.

Once the graph has been updated using the $N$ message passing layers, we collapse it into a 
1-dimensional feature vector using 
\begin{equation}
  \mathbf{y} = \mathcal{F} \left( \left[ \bigoplus_{i \in \mathfrak{F}} \mathbf{n}_i^N, \mathbf{g}
  \right] \right) , \label{eq:last_layer}
\end{equation}
where $\mathcal{F}$ is the last MLP, $\oplus_{i \in \mathfrak{F}}$ the last multi-pooling
operation (done exactly according to Equation \ref{eq:multi-pooling}, but operating over 
all nodes in the graph $\mathfrak{F}$), and $\mathbf{y}$ represents the target 
of the GNN (e.g. $\Omega_{\rm m}$).

All the MLP are constructed by a series of fully connected layers with ReLU activation 
function (except for the last layer, which does not employ an activation function). 
The number of layers, the number of neurons per layer, the weight decay, and the learning rate 
were considered as hyperparameters. The implementation of all the architectures presented in 
this work was done using \textsc{PyTorch Geometric} \citep{pytorch-geometric}.

\subsubsection{Variations of the architecture}
\label{sec:variations}

In Section \ref{sec:where} we investigate whether the information of our model is due to 
clustering, the distribution of velocities, or both. 
For that test, we made use of slightly different architectures to the one outlined above. 
Their main differences are:

\begin{itemize}
 
 \item {\bf Galaxy positions.} 
 
  This model is used to quantify how much information is coming from the clustering of
  galaxies, i.e., it only uses galaxy positions. For that reason, the graphs only contain edge
  features (in the same way outlined above) and no node features. Because of this, the first 
  layer of the model (Equation \ref{eq:edge_model}) operates in a slightly different way:
   \begin{equation}
    \mathbf{e}_{i j}^{(1)} = \mathcal{E}^{(1)} \left( \mathbf{e}_{i j}^{(0)} \right) ,
    \label{eq:edge_model-no_nodes}
   \end{equation}
   \begin{equation}
    \mathbf{n}_{i}^{(1)} = \mathcal{N}^{(1)} \left( \left[  
    \bigoplus_{j \in \mathfrak{N}_i} \mathbf{e}_{i j}^{(1)}, \mathbf{g} \right]  \right) .
    \label{eq:node_model-no_nodes}
   \end{equation}
   Note that other layers operate in exactly the same way as described in Equations \ref{eq:edge_model}-\ref{eq:node_model}.
  
 \item {\bf Galaxy velocities.}

    This model is used to quantify how much information is coming from the distribution of galaxy velocities.
    Therefore, the graphs do not contain any spatial information and we can use {\em deep sets}\footnote{It
    is important to note that, different from a usual neural network (NN), this implementation is invariant
    to permutations (the main property of GNNs) and is made to deliver global information of a structured 
    data.} \citep{Zaheer2017} architectures. In this case, we only have a node model that implements:
   \begin{equation}
    \mathbf{n}_{i}^{(\ell + 1)} = \mathcal{N}^{(\ell + 1)} \left( \mathbf{n}_{i}^{(\ell)}   \right) .
    \label{eq:node_model-DS}
   \end{equation}
   The target quantity is computed using Equation \ref{eq:last_layer}. 
\end{itemize}

\subsection{Likelihood-free inference and the loss function}
\label{sec:free_likelihood_inference}

Our models are trained to infer the value of a given parameter ($\theta_i$, e.g. 
$\Omega_{\rm m}$) by predicting the marginal posterior mean $\mu_i$ and standard 
deviation $\sigma_i$ without making any assumption about the form of the posterior, 
i.e.
\begin{equation}
 \mathbf{y}_i(\mathcal{G}) = [\mu_i(\mathcal{G}), \sigma_i(\mathcal{G})] ,
\end{equation}
where
\begin{align}
 \mu_i ( \mathcal{G} ) & = \int_{ \theta_i } d \theta_i ~ \theta_i ~ p (\theta_i | \mathcal{G} ) \\
 \sigma^2_i ( \mathcal{G} ) & = \int_{ \theta_i } d \theta_i ~ ( \theta_i - \mu_i)^2 ~ p (\theta_i | \mathcal{G} ) \, .
\end{align}
$\mathcal{G}$ represents the input graph and $p (\theta_i | \mathcal{G} )$ is the marginal
posterior, taken according to
\begin{equation}
 p (\theta_i | \mathcal{G} ) = \int_{ \theta_i } d \theta_1 d\theta_2 \dots d \theta_n  
 ~ p (\theta_1, \theta_2, \dots, \theta_n | \mathcal{G} ) .
\end{equation}
In order to achieve this, we made use of a specific loss function following \cite{Jeffrey2020}:
\begin{equation}
 \mathcal{L} = 
 \log \left[ \sum_{ j \in \mathrm{batch} } \left( \theta_{i, j} - \mu_{i, j} \right)^2 \right] + 
 \log \left\{ \sum_{ j \in \mathrm{batch} } \left[ 
 \left( \theta_{i, j} - \mu_{i, j} \right)^2 - \sigma_{i, j}^2 \right]^2 \right\} , \label{eq:loss}
\end{equation}
where $j$ represents the samples in a given batch and $i$ represents the index of the 
considered parameter (e.g. $i=1$ for $\Omega_{\rm m}$). We refer the reader to 
\cite{Paco2022} for the justification of the usage of the logarithms in the above 
expression.

We note that throughout the paper we will be referring to the error of the model as the
quantity described above $\sigma_i$. This error only represents the aleatoric error, and 
therefore does not include the epistemic one, i.e., the error intrinsically related to the 
ML model. 
We have quantified the magnitude of the epistemic errors by training $10$ different models with
the same value of the hyperparameters (the best ones for the considered setup) and calculating
the variance between the predictions of the models. We find that error to be $10\times$ smaller
than the aleatoric one. Therefore, from now on, we will only report aleatoric errors since they 
dominate the total error budget.

\subsection{Training procedure and optimization}
\label{sec:train_optm}

We train our models on graphs constructed from galaxy catalogs of the LH sets of a given 
suite (e.g. the LH set of the Astrid simulations). We initially split the $1000$
LH simulations into training ($850$ simulations), validation ($100$ simulations), and
testing ($50$ simulations). For each simulation, we generate $10$ galaxy catalogs
constructed by taking all galaxies with stellar masses larger than 
$1.3 R \times 10^8~M_\odot/h$, where $R$ is a random number uniformly distributed between 
$1$ and $2$. This strategy is made in order to marginalize over different minimum threshold 
values for stellar masses, as well to increase the number of catalogs used to train the
models\footnote{A similar trick was used in \cite{helen-halos-2022}, where the authors 
employed a similar augmentation in the halo catalogs, choosing them according to a minimum 
number of dark matter particles as a threshold. \cite{helen-prep} also made use of this 
method.}. 
For each catalog, we produce a graph as outlined in Section \ref{sec:the_graph}.

We then train the models using the above architecture for $300$ epochs making use of 
\textsc{Adam} optimizer \citep{Adam} to perform the gradient descent, and a batch size of 
$25$ samples.
The hyperparameter optimization (where we have used the learning rate, the weight decay,
the linking radius, the number of message passing layers, and the number of hidden channels
per layer of the MLPs) was carried out using the \textsc{Optuna} package
\citep{optuna_2019} to perform a Bayesian optimization with Tree Parzen Estimator (TPE) 
\citep{Bergstra2011}. 
We made use of at least $100$ trials to perform this task and we directed \textsc{Optuna}
to minimize the validation loss, computed using an early-stopping scheme, in order to save only 
the model with the minimum validation error. The selected model was used for test subsequently. 

\begin{figure*}
 \centering
 \includegraphics[scale=0.32]{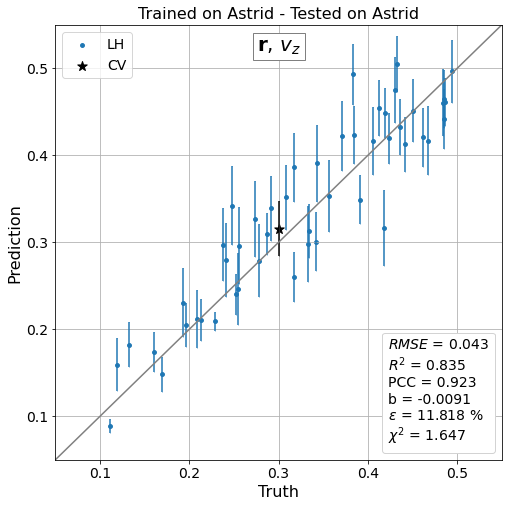}
 \includegraphics[scale=0.32]{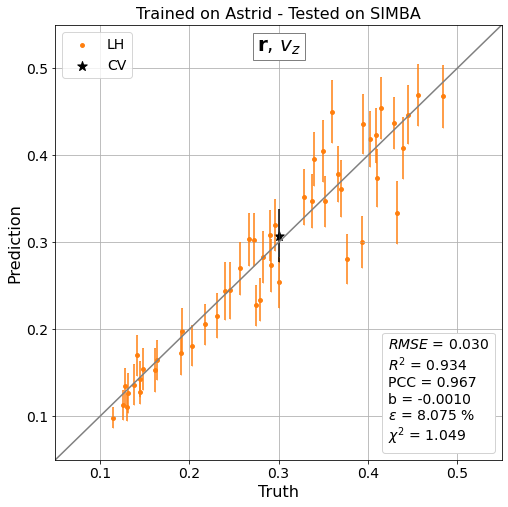}
 \includegraphics[scale=0.32]{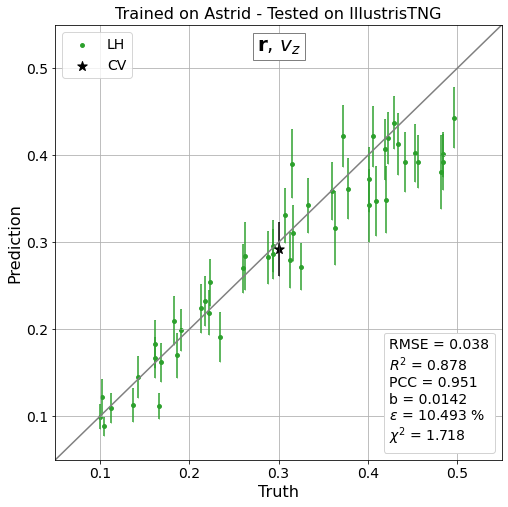}
 \includegraphics[scale=0.32]{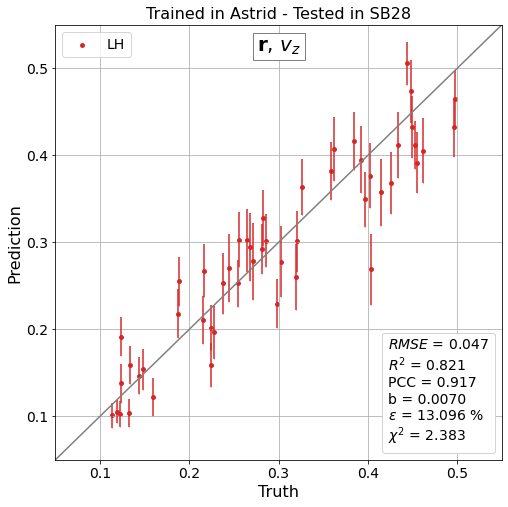}
 \includegraphics[scale=0.32]{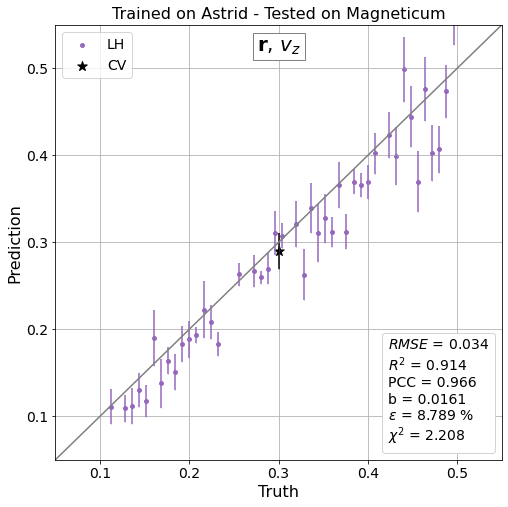}
 \includegraphics[scale=0.45]{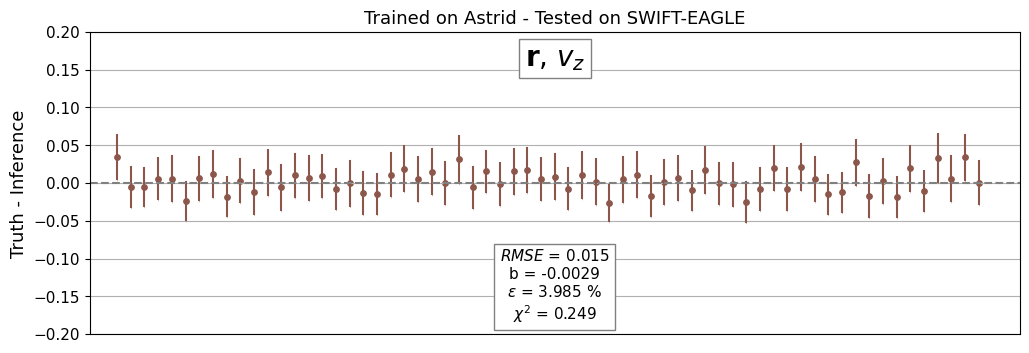}
 \includegraphics[scale=0.45]{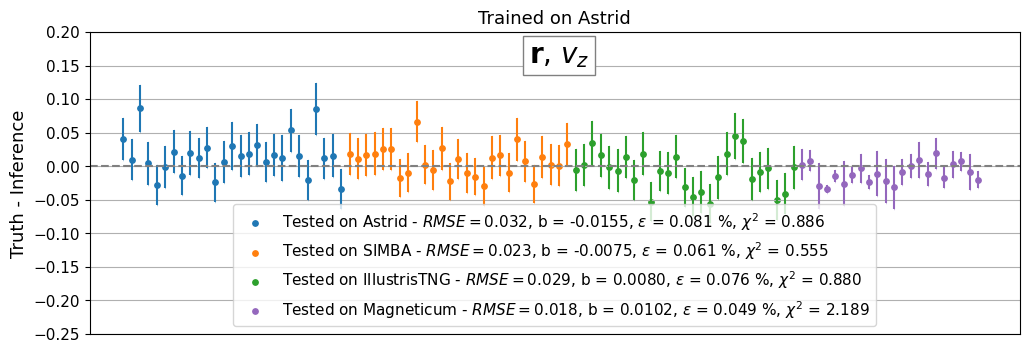}

 \caption{Likelihood-free inference of $\Omega_{\rm m}$ using galaxy {\bf positions} and
 {\bf velocities in the $z$ direction}. We present the results for models {\em trained}
 on {\bf Astrid} and {\em tested} on Astrid (top left), SIMBA (top middle), 
 IllustrisTNG (top right), SB28 (second row left), Magneticum (second row right), and 
 SWIFT-EAGLE (third row). The bottom panel shows the results of testing on CV sets of 
 Astrid, SIMBA, IllustrisTNG, and Magneticum. }
 \label{fig:Astrid}
\end{figure*}

\subsection{Performance Metrics}
\label{sec:scores}

We quantify the accuracy and precision of our models using different metrics that we 
describe below. We consider the true value of the parameter in question for graph $i$ as
$\theta_i$, while we denote as $\mu_i$ and $\sigma_i$ the prediction of the network for the
posterior mean and standard deviation, respectively.

\begin{itemize}

 \item {\bf Root Mean Squared Error (RMSE):}
  \begin{equation}
   \text{RMSE} = \sqrt{ \frac{1}{N} \sum_{i = 1}^N 
   \left( \theta_i - \mu_i \right)^2 } . \label{eq:RMSE}
  \end{equation}
Low values of the RMSE indicate the model is precise.

 \item {\bf Coefficient of determination:}
  \begin{equation}
    R^2 = 1 - \frac{ \sum_{i = 1}^N \left( \theta_i - \mu_i \right)^2 }{ 
    \sum_{i = 1}^N \left( \theta_i - \bar{\theta}_i \right)^2 } , \label{eq:R2} 
  \end{equation}
  where $\bar{\theta}_i = \frac{1}{N} \sum_{i = 1}^N \theta_i$. Values close to $1$ indicate 
  the model is accurate.

 \item {\bf Pearson Correlation Coefficient (PCC):}
  \begin{equation}
   \label{eq:PCC}
   \rm{PCC} = \frac{\rm{cov} \left( \theta, \mu \right)}
       {\sigma_\theta \sigma_\mu} .
  \end{equation}
  This statistic measures the positive/negative linear relationship between truth values 
  and inferences: good values are close to $\pm 1$, and worse closer to $0$. 
  It gives an idea of the accuracy of the model.

 \item {\bf Bias:}
  \begin{equation}
   b = \frac{1}{N} \sum_{i = 1}^N \left( \theta_i - \mu_i \right) .
  \end{equation}
This statistic quantifies how much the inferences are ``biased'' with respect to the truth 
values; better values are close to $0$.
 
 \item {\bf Mean relative error:}
  \begin{equation}
   \epsilon = \frac{1}{N} \sum_{i = 1}^N \frac{ | \theta_i - \mu_i | }{ \mu_i } .
  \end{equation}
Low values of this statistic indicate the model is precise.
  
 \item {\bf Reduced chi squared:}
  \begin{equation}
   \chi^2 = \frac{1}{N} \sum_{i = 1}^N \left(\frac{\theta_i - \mu_i }{\sigma_i} \right)^2  .  
  \end{equation}
  This statistic quantifies the accuracy of the estimated errors.
  Values of $\chi^2$ close to $1$ indicate the magnitude of the errors (posterior standard 
  deviation in our case) is properly inferred, while values larger/smaller than $1$ indicate
  the model is under/over predicting the errors.
  
\end{itemize}
We make use of these statistics to quantify the accuracy, precision, and bias of a given 
model in the test set. Note that in some cases we omit to report the value of some of these 
statistics for clarity, or when the statistics are not well defined (e.g. when tested on the 
CV set).

\section{Results} 
\label{sec:results}

In this section we present the main results of testing our GNN models on galaxy catalogs
with different cosmologies, astrophysical parameters, and subgrid physics models from the
catalogs used for training. We start by showing the results of our best model, which only needs 
$3$D galaxy positions and $1$D velocity components, in Section \ref{sec:Astrid-best_model}. 
We then attempt to increase the precision of the model by adding more galaxy properties, 
particularly stellar mass, in Section \ref{sec:Astrid-SM}. Next, we investigate the origin of 
the information extracted by our models in Section \ref{sec:where}. 

Note that we focus our analysis entirely on $\Omega_{\rm m}$. This is because our 
constraints on $\sigma_8$ are very weak.
We provide further details on this in Appendix \ref{sec:sigma8}. All results below are 
shown for catalogs built with galaxies with a minimum value of stellar mass as 
$M_{\star} = 1.95 \cdot 10^8 ~M_\odot/h$, a value right in the middle of the threshold used in 
our training criteria\footnote{We have checked that our results are not very sensitive to the 
particular stellar mass cut we take, as long as we are not very close to the training 
boundaries.}.

\begin{figure*}
 \centering
 \includegraphics[scale=0.32]{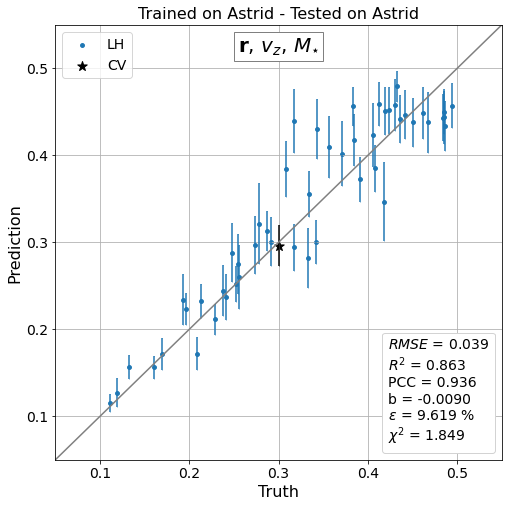}
 \includegraphics[scale=0.32]{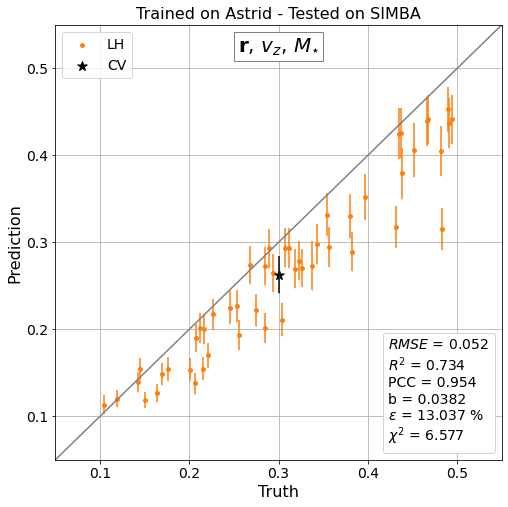}
 \includegraphics[scale=0.32]{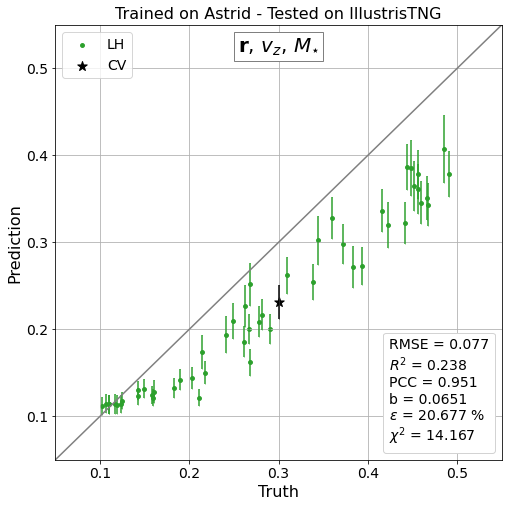}
 \includegraphics[scale=0.32]{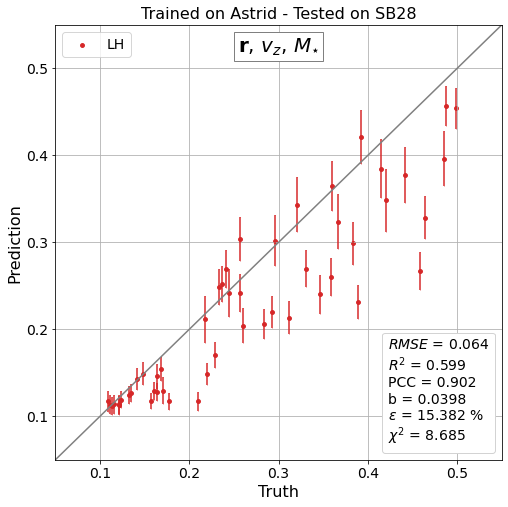}
 \includegraphics[scale=0.32]{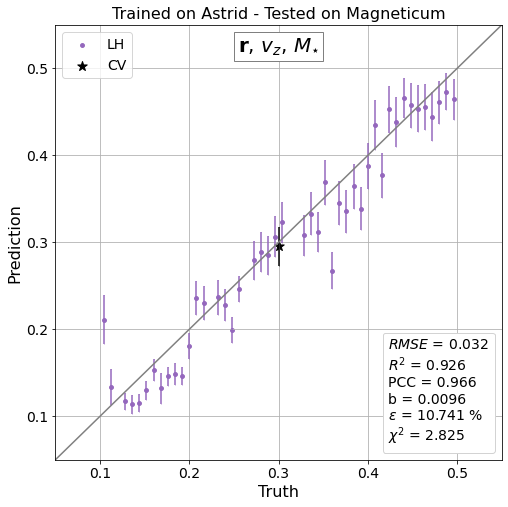}
 \includegraphics[scale=0.45]{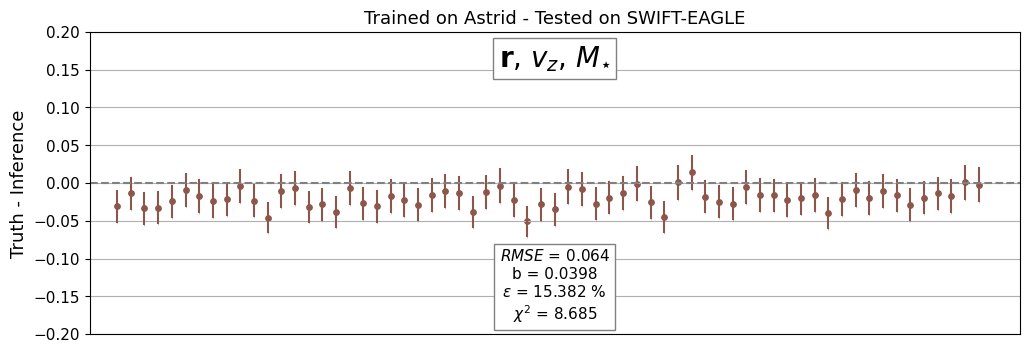}
 \includegraphics[scale=0.45]{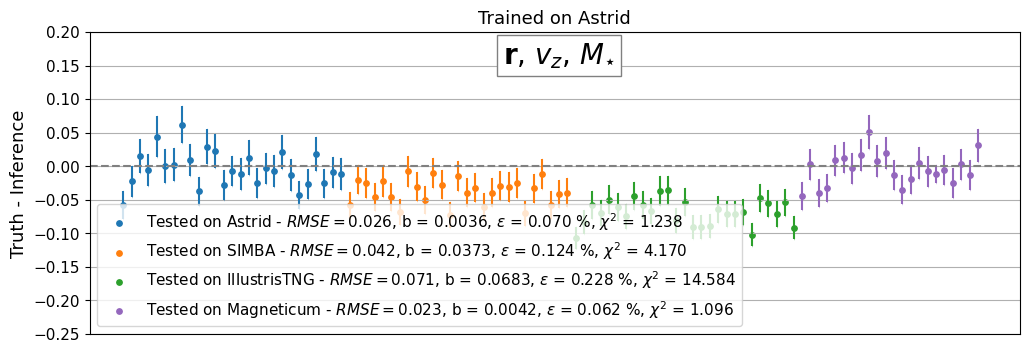}
 
 \caption{Likelihood-free inference of $\Omega_{\rm m}$ using galaxy {\bf positions}, {\bf 
 velocities in the $z$ direction}, and {\bf stellar mass}. We present the results for a model 
 {\em trained} on {\bf Astrid} and {\em tested} on Astrid (top left), SIMBA (top middle), 
 IllustrisTNG (top right), SB28 (second row left), Magneticum (second row right), and 
 SWIFT-EAGLE (third row). The bottom panel shows the results of testing on CV sets of Astrid, 
 SIMBA, IllustrisTNG, and Magneticum.}
 \label{fig:Astrid-SM}
\end{figure*}

\subsection{Positions \& velocities} 
\label{sec:Astrid-best_model}

We start by showing the results of training GNNs on catalogs that only contain the positions
and velocities (only the $z$ component)\footnote{Due to homogeneity and isotropy, the results 
presented choosing the $z$ component of the velocity are equivalent to choosing either $x$ or 
$y$ ones.} of galaxies to infer the value of $\Omega_{\rm m}$. 
We have trained models using galaxy catalogs from the LH sets of the Astrid, IllustrisTNG,
and SIMBA simulations. We then test these models on all other galaxy catalogs not included in 
their training set.

We found that the model trained on Astrid galaxy catalogs exhibits the best 
extrapolation properties, so we focus our analysis on it. The success of the model 
trained on Astrid can be associated with (1) the variety in the number of galaxies along
the Astrid catalogs in LH sets, which vary from small to large numbers of galaxies 
($N \in [30, 5,000]$ -- see more details in the Appendix \ref{sec:graph_details}) and 
(2) Astrid produces larger variations in some galaxy properties given the parameter variations 
in the LH set \citep{Yueying-prep}. 
We show the results of the models trained on IllustrisTNG and SIMBA catalogs in 
Appendix \ref{sec:SIMBA_IllustrisTNG}.
In addition, we trained a model on SB28 set, but even so, the model does not show good predictions 
when tested on the other simulations.

In Figure \ref{fig:Astrid} we show the results of testing the model on galaxy catalogs from 
the LH sets of Astrid (top left), SIMBA (top middle), IllustrisTNG (top  right), SB28 (second 
row left), Magneticum (second row right), and SWIFT-EAGLE (third row). In all these plots 
(apart from SB28 and SWIFT-EAGLE) we present the average (of their 
mean and standard deviation) across all of their CV boxes as a black point at 
$\Omega_{\rm m}=0.3$.
The results of testing the model on galaxies catalogs from the CV set of the different suites 
are shown in the bottom panel. 
Note that for clarity we only show $50$ randomly selected samples of the predictions for all 
the LH results\footnote{In the case of Astrid we only have $50$ samples in the test set since 
the majority of the LH set was used for training.}. 
We stress that even if we only show the results for 50 random catalogs, the numbers reported 
for the different performance metrics (e.g. RMSE) are evaluated using all catalogs in the test 
set (e.g. 1000 catalogs for IllustrisTNG).

When using the model trained on Astrid and testing it on itself, we find that the GNN is able
to infer $\Omega_{\rm m}$ with $RMSE = 0.043$, $R^2 = 0.835$, 
$PCC = 0.923$, $b = - 0.0091$, $\epsilon = 11.8 \%$, and $\chi^2 = 1.647$. 
These numbers indicate the model is accurate, precise, unbiased, and its errors are 
only slightly under predicted\footnote{To show the scores for the best model while testing it 
on Astrid and Magneticum, we removed respectively one and four predictions that correspond to a 
$\chi^2$ larger than $14.0$. They are points in the test set that achieved this bad inference 
and that we call ``outliers''. Outliers not only because of the bad scores but mainly 
because they correspond to particular realizations in the LH set with extreme values for the 
astrophysical parameters, which are realizations far away from the fiducial model. We do not 
follow this procedure in the other models (apart from the best model, trained on Astrid using 
only galaxy positions and $z$ component of the velocity) because they end up with a huge number 
of ``bad'' predictions, not only in the matter of fact to this issue.}. 
While testing that model on the other simulations the performance
metrics are in the ranges: $RMSE \in [0.015, 0.047]$, $R^2 \in [0.821, 0.934]$, 
$PCC \in [0.917, 0.967]$, $b \in [- 0.0010, 0.0161]$, $\epsilon \in [4.0, 13.1] \%$, and 
$\chi^2 \in [0.249, 2.383]$, showing that the model extrapolates very well, as can also be
seen in Figure \ref{fig:Astrid}. Note that the model performs best on SIMBA and SWIFT-EAGLE, 
and worst on SB28. This indicates that, while the model is generally robust, even when tested 
on SB28, it becomes increasingly difficult to extrapolate predictions over distant regions in 
parameter space.

We have included a test using IllustrisTNG300 box in order to test the importance of 
super-sample covariance effects. Basically, the lack of power on scales larger than our boxes 
can affect both the abundance and clustering of galaxies 
\citep{Hu2003, Hamilton2006, Takada2007, Li2014}. We find that our method can partially 
account for these effects. We provide further details in Appendix \ref{sec:TNG300}.

We now discuss the performance of the model on galaxy catalogs from the CV set. 
We find that our model works better when tested on 
the CV catalogs compared to the LH and SB sets. This could be due to the fact that the 
cosmology and astrophysics of those models lie exactly in the center of the training set. 
Those configurations are less prone to biased results, although it is interesting to observe 
that cosmic variance effects are not the main contribution to the error budget.
Finally, all the different simulations end up with differences lower than 
$5 \%$ (apart from some boxes of Astrid or SIMBA, where we achieve differences 
\{truth - inference\} up to $10 \%$) for the best model, once again being accurate, precise, 
and without any bias. 

We conclude this part by emphasizing the overall good accuracy of our model, which accounts 
for cosmic variance, marginalizes over astrophysics, and is robust to changes in halo/subhalo 
finder and subgrid physics models.
On top of this, the fact that the model works so well even in full extrapolation mode (e.g. 
when being tested on the SB28 simulations) indicates that the network may have learned 
physical relations 
(coming from the galaxies phase-space distribution)
rather than a common feature among simulations.

\subsection{Positions, velocities, \& stellar masses} 
\label{sec:Astrid-SM}

We now investigate whether we can make our model more precise, while keeping it robust, by 
considering an additional galaxy property: the stellar mass. For this, we construct graphs 
in the standard way (as described in Section \ref{sec:methodology}) but taking as node features 
both velocity and stellar mass: $[v_z, M_{\star}]$ (properties normalized as described 
in Section \ref{sec:the_graph}). 
We then train GNN models using catalogs from the Astrid LH set.

We present the results in Figure \ref{fig:Astrid-SM}. When testing the model on galaxy 
catalogs from the Astrid LH set we find that the results improved for almost all the metrics: 
$RMSE = 0.039$, $R^2 = 0.863$, $PCC = 0.936$, $b = - 0.0090$, $\epsilon = 9.62 \%$, and
$\chi^2 = 1.849$, which means that the GNN was able to extract more information from 
the catalogs. On the other hand, when testing the model on the galaxy catalogs from 
the other simulation suites the scores worsen: 
$RMSE \in [0.032, 0.077]$, $R^2 \in [0.238, 0.926]$, $PCC \in [0.902, 0.966]$, 
$b \in [0.0096, 0.0651]$, $\epsilon \in [10.7, 20.7] \%$, and $\chi^2 \in [2.825, 14.167]$. 
In other words, the model has become more precise when tested on itself, at the expense of
becoming less accurate, when tested on other simulation sets. 
It is worth noting that some metrics actually improved when tested on galaxy 
catalogs from Magneticum, as seen in Figure \ref{fig:Astrid-SM}. 
It is not clear to us what could be the explanation behind this: whether it is either 
a coincidence or due to the fact that galaxies in Astrid and Magneticum are more alike somehow 
while considering this specific galaxy property.

Our results are in agreement with those of \cite{pablo-galaxies-2022} who performed a 
similar analysis with galaxy catalogs whose node features were the maximum circular 
velocity, the stellar mass, the galaxy radius, and the star metallicity. While the model of
those authors was more precise than ours (likely due to the use of additional galaxy
properties), it was not robust. However, our models are slightly more robust; we believe
this could be due to the fact that we use catalogs with different stellar mass thresholds to
train the models, which overcomes the differences due to the fact that we are marginalizing 
over different stellar mass thresholds. This conclusion agrees with 
what \cite{helen-halos-2022} have found using the same idea of marginalization over an 
augmentation technique.

We reach similar conclusions when testing our models on galaxy catalogs from simulations 
of the CV sets (see the last panel of Figure \ref{fig:Astrid-SM}), especially noticing that
we have obtained a bias in the predictions for the different simulations. We emphasize the
importance of testing the models on simulations as diverse as possible. Should we only have
galaxy catalogs from Astrid and Magneticum simulations, we could reach the wrong conclusion
that the model was both more precise and accurate than the one constructed using only
positions and velocities.

\begin{figure*}
 \centering
 \includegraphics[scale=0.32]{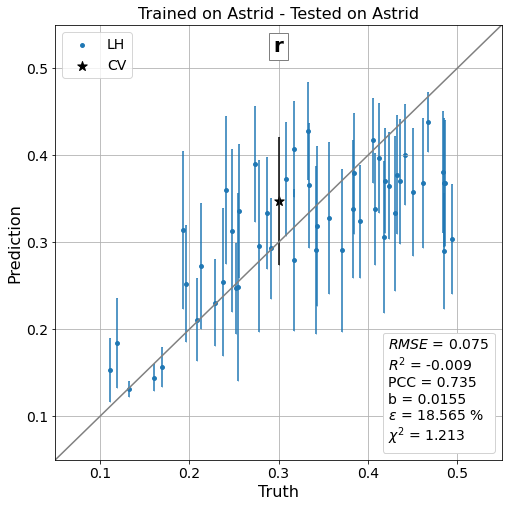}
 \includegraphics[scale=0.32]{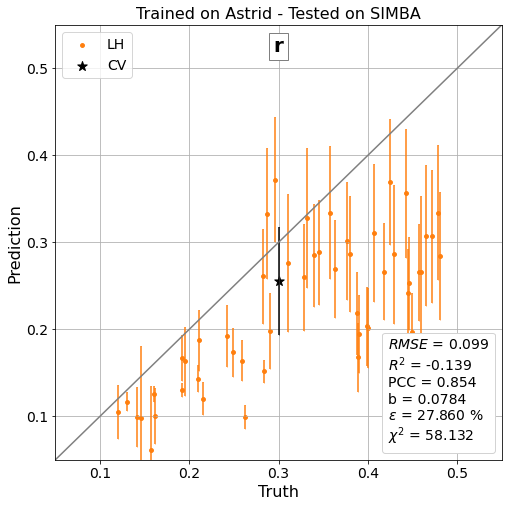}
 \includegraphics[scale=0.32]{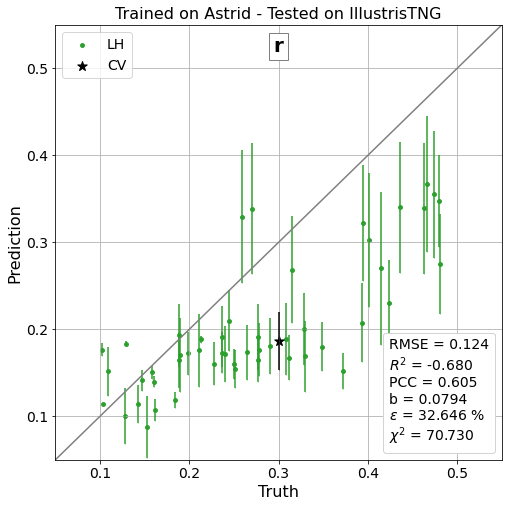}
 \includegraphics[scale=0.32]{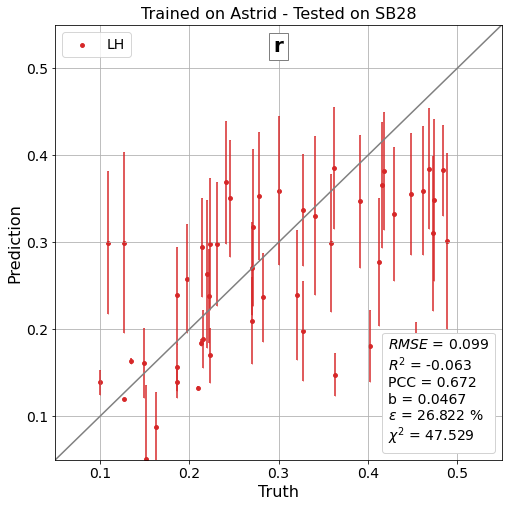}
 \includegraphics[scale=0.32]{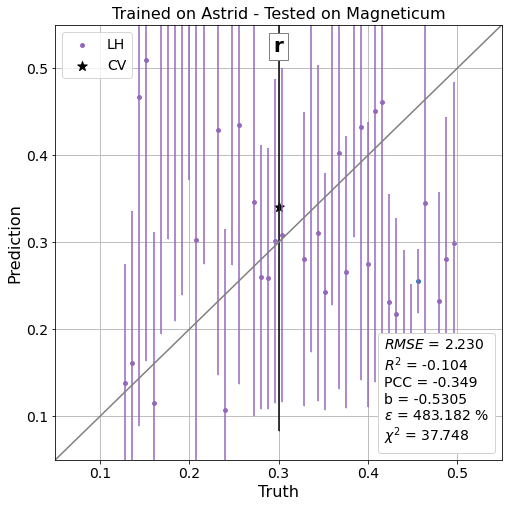}
 \includegraphics[scale=0.45]{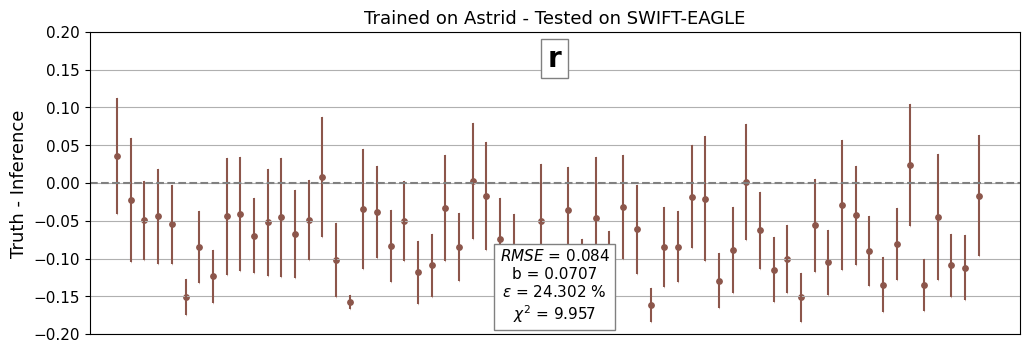}
 \includegraphics[scale=0.45]{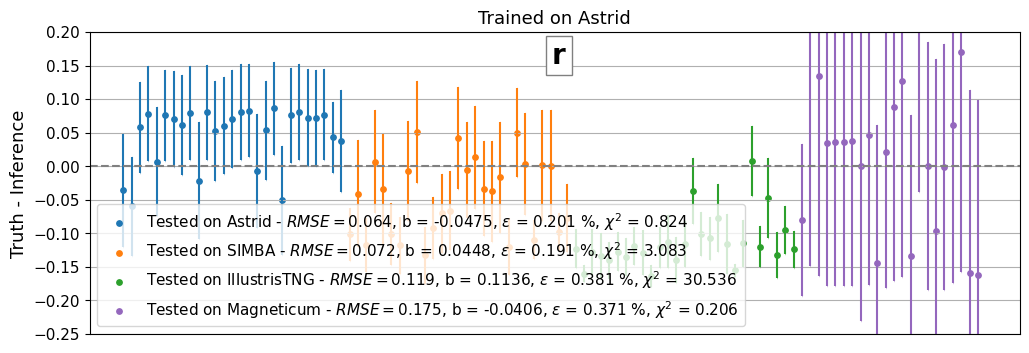}
 \caption{Likelihood-free inference of $\Omega_{\rm m}$ using {\bf only} galaxy 
 {\bf positions}. 
 We present the results for models {\em trained} on {\bf Astrid} and {\em tested} on 
 Astrid (top left), SIMBA (top middle), IllustrisTNG (top right), SB28 (second row left), 
 Magneticum (second row right), and SWIFT-EAGLE (third row). The bottom panel shows the results 
 of testing on CV sets of Astrid, SIMBA, IllustrisTNG, and Magneticum.}
 \label{fig:Astrid-OP}
\end{figure*}

\begin{figure*}
 \centering
 \includegraphics[scale=0.32]{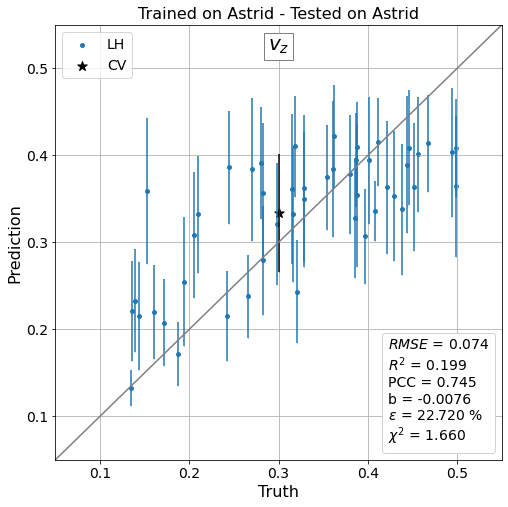}
 \includegraphics[scale=0.32]{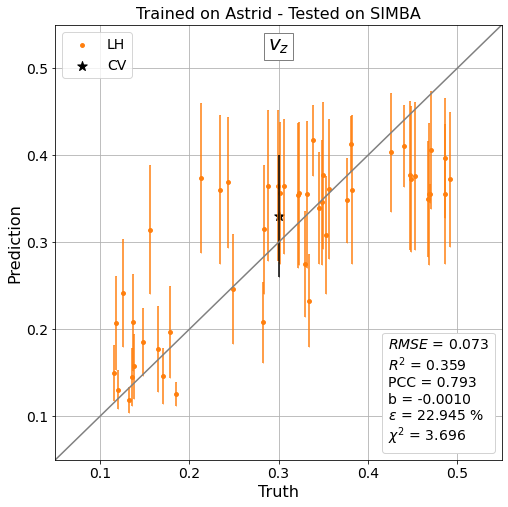}
 \includegraphics[scale=0.32]{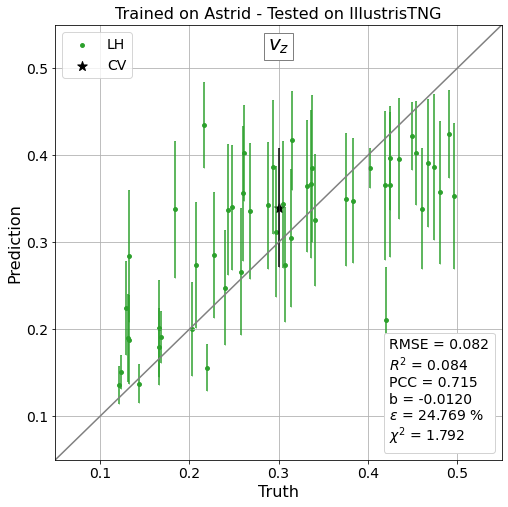}
 \includegraphics[scale=0.32]{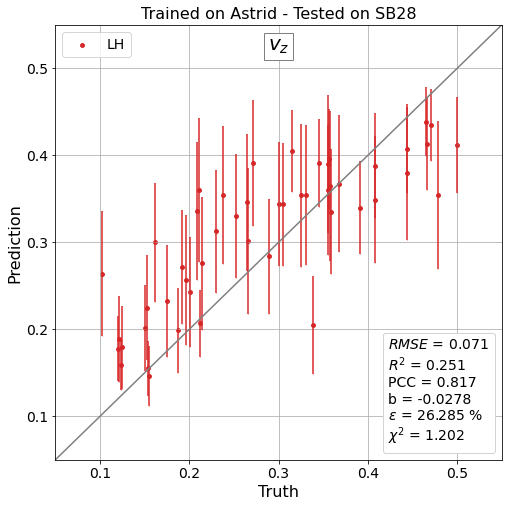}
 \includegraphics[scale=0.32]{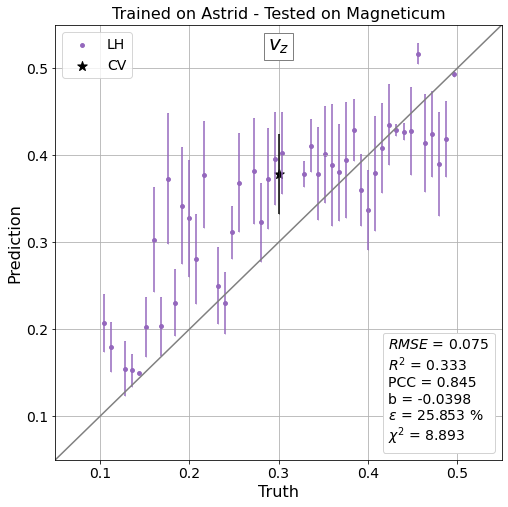}
 \includegraphics[scale=0.45]{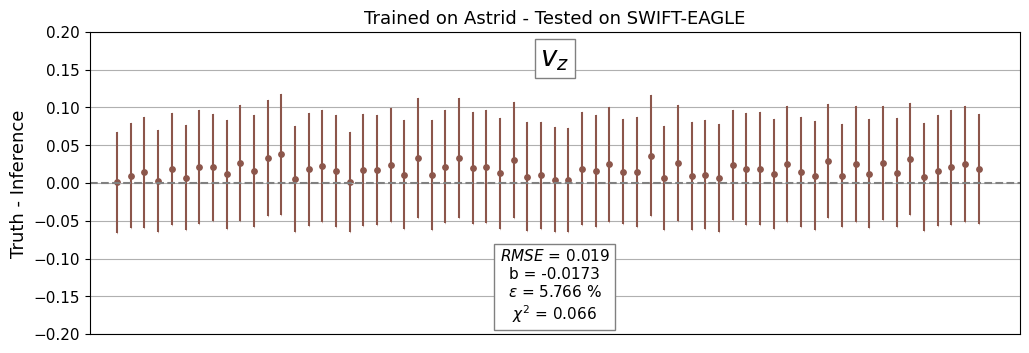}
 \includegraphics[scale=0.45]{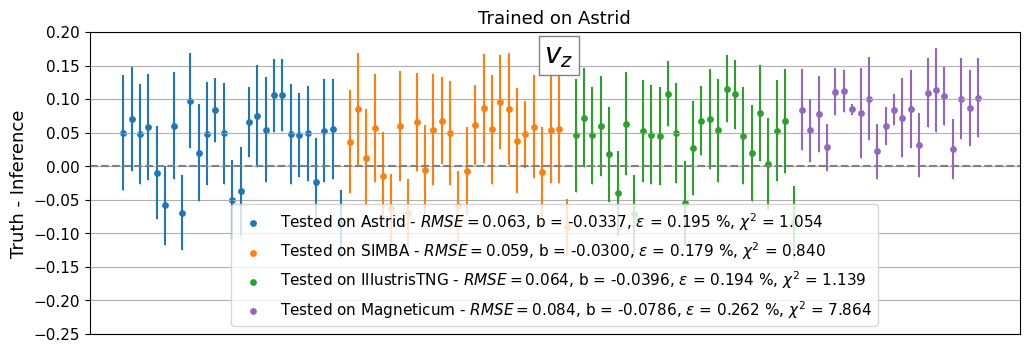}
\caption{Likelihood-free inference of $\Omega_{\rm m}$ using galaxy {\bf velocities in $z$ 
direction}. We present the results for a model {\em trained} on {\bf Astrid} and 
{\em tested} on Astrid (top left), SIMBA (top middle), IllustrisTNG (top right), SB28 
(second row left), Magneticum (second row right), and SWIFT-EAGLE (third row). The bottom panel 
shows the results of testing on CV sets of Astrid, SIMBA, IllustrisTNG, and Magneticum.}
 \label{fig:Astrid-DS}
\end{figure*}

\subsection{Where does the information come from?}
\label{sec:where}

We now investigate where the information from our robust model (discussed in Section 
\ref{sec:Astrid-best_model}) comes from. 
Since in that model we only made use of galaxy positions and velocities, there are only three 
possibilities: 1) the information is coming from the positions of galaxies (clustering), 2) 
the information is coming from the distribution of galaxy velocities, and 3) the information 
is coming from both positions and velocities.
Note that we are not considering attributing the importance to the level of information 
coming from the number of galaxies in the catalogs because: a) as mentioned in the Footnote \ref{g_number}, 
this global property only improved slightly the results, and b) we do not have a considerable number of
catalogs with the same number, or even with the same range of the number, of galaxies (see Appendix 
\ref{sec:graph_details}). This last reason should result in worse predictions due to the lack of data to train the
machinery and would not allow to test it in all the different sub-grid physics models (which is the case of 
Magneticum, which only contains boxes with thousands of galaxies - see again Appendix \ref{sec:graph_details}).

In order to address the first possibility we have made use of graphs where the nodes do not 
contain any property. We train the model, in the Astrid suite, using the first slightly 
different GNN architecture described in Section \ref{sec:variations}: galaxy positions, i.e., 
using the prescription presented in Equations 
\ref{eq:edge_model-no_nodes}-\ref{eq:node_model-no_nodes} for 
the first message passing layer. We then test the model on the different graphs from 
different simulation suites. The results are presented in Figure \ref{fig:Astrid-OP}, 
following the same scheme as Figure \ref{fig:Astrid}. In all the tests the results are 
visibly worse (with large error bars) and significantly biased (when extrapolating to the other 
simulations). More specifically, we found: $RMSE \in [0.084, 2.230]$, 
$R^2 \in - [0.680, 0.063]$, $PCC \in [- 0.349, 0.854]$, $b \in [- 0.5305, 0.0467]$, 
$\epsilon \in [24.3, 483.2] \%$, and $\chi^2 \in [9.957, 70.730]$. 
While testing the model in the CV sets we found a low performance for all the 
metrics analyzed, with larger error bars. 
Our results are qualitatively in agreement with those of \cite{pablo-galaxies-2022}, who 
performed a similar analysis but with galaxy catalogs with a fixed stellar mass threshold and 
did not use Astrid as the training set. 
From this test, we conclude that the network cannot be extracting the information just from 
galaxy clustering.

Next, we train a deep set model (see the second model presented in Section 
\ref{sec:variations}: 
galaxy velocities) on galaxy catalogs that only contain the $z$ component of the galaxy
velocities (i.e. there are no galaxy positions) and, then, we employed Equation
\ref{eq:node_model-DS}. We used Astrid simulation to train the model.
Figure \ref{fig:Astrid-DS} displays the results. Also in this case we find that the model 
performs poorly: 
$RMSE \in [0.019, 0.082]$, $R^2 \in [0.084, 0.359]$, $PCC \in [0.715, 0.845]$, 
$b \in - [0.0398, 0.0010]$, $\epsilon \in [5.8, 26.3] \%$, and $\chi^2 \in [0.066, 8.893]$.
These results are distinct from what \cite{pablo-galaxies-2022} found (while using a deep 
set as well), whose scores were comparable to the ones from the GNN. Note that those authors 
used more galaxy properties and we only use the 1D velocity component.
The results for catalogs of the CV sets have large error bars and poor values for all the 
metrics. We then conclude that galaxy velocities can not be alone the origin of the 
information extracted by the network. 

The above tests indicate that the network is making use of both positions and 
velocities to infer the value of $\Omega_{\rm m}$. Another important point to highlight is 
that the models trained on galaxy positions alone and galaxy velocities alone, although not 
very precise, seem to also not be robust. This may indicate that the model that uses 
galaxy positions and velocities may be extracting robust information due to constraints in 
phase space (e.g. the necessity to fulfill the continuity equation), directly encoding 
effective information on $\Omega_{\rm m}$. 

These findings may be related to some previous ideas that correlate with the matter 
content of the Universe to galaxy positions and peculiar velocities 
\citep{peebles1980, kaiser, Cen1994, Strauss1995}, and that motivated a number of efforts
towards peculiar velocity surveys \citep{Howlett2017, Kourkchi2020, SLOAN2022}. 
Besides, our results agree with the findings of \cite{helen-halos-2022} who used GNNs to 
predict $\Omega_{\rm m}$ based on positions and velocity modulus, but for DM halo catalogs. 
Moreover, this intricate relation motivates a deep analysis of the direct interpretation of the
network predictions, which is being taken into account by us in \cite{helen-prep}, combining 
symbolic regression with the GNNs.

\section{Discussion and conclusions} 
\label{sec:conclusions}

The quest to extract the maximum information from galaxy redshift surveys has motivated the 
development of many different approaches \citep{Efron1980, FKP, Taylor2013, MTPK, Alan2018,
Hahn2020, Uhlemann2020, Gualdi2021, Banerjee2021, deSanti2022JCAP, CARPool2022} and the
upcoming data from the current and next generation of surveys \citep{SKA1999, Laureijs2011,
Ellis2012, Amendola2012, Benitez2014, Roman2015, DESI, Euclid2022-Tiago_Castro, JWST}, is 
pressing this field of research. While we do not have a final answer to this question ML
techniques are appearing as a promising tool to tackle this problem 
\citep{Ravanbakhsh2017, Ntampaka2020, Mangena2020, Hassan2020, Paco2021, Cole2021, lucia2022}. 
In particular, GNNs stand out as good machinery to extract 
cosmological information from galaxy and halo catalogs from simulations 
\citep{pablo-galaxies-2022, helen-halos-2022, lucas2022, Anagnostidis_2022}.

GNNs are ideal methods to analyze galaxy redshift surveys because: 1) they are designed to 
work with sparse and irregular data \citep{Gilmer2017, Battaglia2018, Bronstein2021}; 2) it 
is easy to construct models that fulfill physical symmetries \citep{pablo-galaxies-2022}; 
3) they do not apply any cutoff on the scale to extract information. 
Perhaps the most challenging task associated with ML methods is their robustness 
\citep{robustness-Hassani2022}, a hard question already explored using 2D maps with CNNs
\citep{Paco2021}, tabular data \citep{one_gal-2022}, and galaxy catalogs 
\citep{pablo-galaxies-2022}. 
The reason behind the lack of robustness of the models is unclear and can be due to multiple 
factors: 
1) data sets do not overlap;
2) models may be learning no physical effects (e.g. numerical artifacts);
3) data representation is different. 
We emphasize that precision is completely irrelevant without accuracy. 
The only way to deploy ML models to perform analysis with real data is to employ accurate 
models. Thus, robustness lies at the heart of this problem.

In this work, we have trained GNN models on thousands of galaxy catalogs from 
state-of-the-art hydrodynamic simulations of the CAMELS project to infer the value of 
$\Omega_{\rm m}$ at the field-level using a likelihood-free approach. More importantly, we
have investigated the robustness of the models by testing them on galaxy catalogs from 
simulations run with completely different codes to the ones used for training. We now 
outline the main takeaways from this work:

\begin{itemize}

\item The model trained on Astrid catalogs that only contain galaxy positions and velocities 
(the $z$ component) is able to infer the value of $\Omega_{\rm m}$ with $\sim12\%$ precision 
and accuracy when tested on Astrid catalogs with different cosmologies and astrophysical 
parameters. 

\item The performance is similar when tested on galaxy catalogs from other galaxy formation 
simulations (each with different cosmology and astrophysics) run with four different 
hydrodynamic codes: IllustrisTNG, SIMBA, Magneticum, and SWIFT-EAGLE. This fact illustrates the 
robustness of the model under variations of the underlying subgrid physics.

\item It also works well when tested on the SB28 set of the IllustrisTNG suite: a collection of 
$1024$ simulations that varies $28$ parameters ($5$ cosmological and $23$ astrophysical) and 
therefore goes well beyond the diversity used to train the model (where only 6 parameters are 
varied). 

\item Our model is also robust to changes in the halo/subhalo finder: the galaxy catalogs of 
the SWIFT-EAGLE simulations were constructed employing \textsc{VELOCIraptor}, a different 
method than the one used for training (\textsc{SubFind}). When we tested our model on 
SWIFT-EAGLE catalogs we still obtained good predictions.

\item The above constraints were obtained using a very small volume $(25~h^{-1}{\rm Mpc})^3$ 
that only contains $\sim1000$ galaxies with stellar masses above $\sim2\times10^8~M_\odot/h$ 
at $z=0$. We note that some galaxy catalogs contain a much larger ($\sim 5,000$, which is the 
case of Magneticum simulations) or smaller ($\sim 30$, in some Astrid boxes) number of 
galaxies and the model still performs well on those.

\item When training our models on galaxy catalogs that contain positions, velocities, and 
stellar masses we are able to build models that are more precise but less accurate. In fact, 
those models are no longer robust across different simulation codes and therefore could not be 
used with real data.

\item We find that our models are extracting information from both galaxy positions and 
velocities. Furthermore, models trained using catalogs that only contain galaxy positions or 
galaxy velocities  are not only less precise but also less accurate. We speculate that having 
both positions and velocities may improve the accuracy of the models as the phase-space 
distribution is constrained by physical arguments, such as the continuity equation, that need 
to be fulfilled independently of cosmology, astrophysics, and subgrid model employed.

\end{itemize}

Given the precision and accuracy of our model, it will be interesting applying it to peculiar 
velocity surveys such as the SLOAN catalog \citep{SLOAN_catalog-2022} or even the Cosmicflows-4 
catalog \citep{Kourkchi2020}. We note that several steps need to be carried out before 
performing such a task:

\begin{itemize}

 \item The method needs to be shown robust to changes in super-sample covariance. This is 
 because in this analysis we did not account for such effect at the training stage. If the 
 method is not robust to this effect, we should retrain our models on galaxy catalogs from 
 larger volumes or catalogs that include the super-sample covariance effect. We note that 
 preliminary work indicates that the models can deal with this effect, at least partially. We 
 refer the reader to Appendix \ref{sec:TNG300}) for further details.

 \item Through this work we are dealing with peculiar velocities from simulations.
 Therefore, we do not take into account any model or changes to consider observational errors in this quantity.
 The peculiar velocities of galaxies cannot be measured with infinite precision. 
 A quantification of how the error on the peculiar velocities propagates into the constrain in 
 $\Omega_{\rm m}$ needs to be performed.

\item An investigation on whether selection effects may affect the results is also needed, as 
some surveys rely on particular tracers (e.g. supernovae) that are not available on all 
galaxies above a certain stellar mass, as we consider here. Moreover, we plan to
investigate the effect of increasing the number of galaxies and decreasing the number density in
future work, competitive effects that will arise in real data and that are respectively low and high in
the present analysis.

\end{itemize}

The possible application of this machinery to real data relies on one inherent limitation
of the presented methodology, a question that is still related to robustness. This is because 
the GNN will be able to extrapolate their predictions only if applied to something compatible 
with the data set on which it has been trained. In other words, if the CAMELS suite of 
simulations will be able to capture the main characteristics of our observable Universe.
We plan to carry out these tasks in future work and show if the model can be robust, 
accurate, and precise to be able to infer a good estimate for $\Omega_{\rm m}$.\\ 

We now discuss the similarities and differences between this paper and previous works:

\begin{itemize}

 \item \cite{pablo-galaxies-2022}: our best model was achieved using a new CAMELS 
 hydrodynamical set of simulations: the Astrid catalogs (different of IllustrisTNG and SIMBA 
 suites, used in that work); only 3D galaxy positions and 1D velocity components carried all 
 the information (differently from what these authors have considered, using stellar metallicity, 
 galaxy radius, and maximum circular velocity too); we employed in the training stage a 
 marginalization over different minimum values for stellar mass thresholds, instead of 
 considering only one for all the ML stages; therefore, our results are robust over different 
 subgrid physics, what does not happen in that work.

 \item \cite{helen-halos-2022}: we have trained our models on galaxy, rather than of halo 
 catalogs from N-body simulations; we make use of galaxy observables as input information 
 (positions and velocity in only one direction), different from considering the modulus of 
 galaxy peculiar velocities.
 
 \item \cite{lucas2022}: these authors used the Quijote halo suite \citep{Quijote_sims}, which 
 does not consider subhalo properties. Besides, their analysis utilizes a different method to 
 compute posteriors than the one employed here. 
 
 \item \cite{Anagnostidis_2022}: in all our analyses we make use of hydrodynamical catalogs, 
 considering astrophysics information. These authors consider lightcones from halo catalogs 
 which is not taken into account here.

\end{itemize}

Overall, this paper presents a new method to study cosmology using the clustering and velocities of
galaxies at the field-level, without imposing any cut on scale, that seems robust to changes in
cosmology, astrophysics, subgrid physics, and galaxy identification algorithms. 



\section*{Acknowledgments}


We would like to thank David Spergel, Ravi K. Sheth, Michael Strauss, 
Tamara Davis, Natália V. N. Rodrigues, Joop Schaye, Matthieu Schaller, Lucia A. Perez,
and the CAMELS team for the enlightening discussions and valuable comments. 
We thank the São Paulo Research Foundation (FAPESP), the Brazilian National Council for
Scientific and Technological Development (CNPq), and the Simons Foundation for financial
support. 
NSMS acknowledges financial support from FAPESP, grants
\href{https://bv.fapesp.br/en/bolsas/187647/cosmological-covariance-matrices-and-machine-learning-methods/}{2019/13108-0} 
and \href{https://bv.fapesp.br/en/bolsas/202438/machine-learning-methods-for-extracting-cosmological-information/}{2022/03589-4}. 
The CAMELS project is supported by the Simons Foundation and NSF grant AST 2108078. TC is supported by the INFN
INDARK PD51 grant and the FARE MIUR grant ``ClustersXEuclid'' R165SBKTMA.
EH acknowledges supported by the grant agreements ANR-21-CE31-0019 / 490702358 from the French 
Agence Nationale de la Recherche / DFG for the LOCALIZATION project.
DAA acknowledges support by NSF grants AST-2009687 and AST-2108944, CXO grant TM2-23006X, and 
Simons Foundation award CCA-1018464.
The research in this paper made use of the SWIFT open-source simulation code 
\citep[\url{http://www.swiftsim.com},][]{Schaller2018} version 1.2.0.
The training of the GNNs has been carried out using  graphics processing units (GPUs) from
Simons Foundation, Flatiron Institute, Center of Computational Astrophysics.


%






\appendix

\section{Graph details}
\label{sec:graph_details}

\begin{figure*}
 \centering
 \includegraphics[scale=0.82]{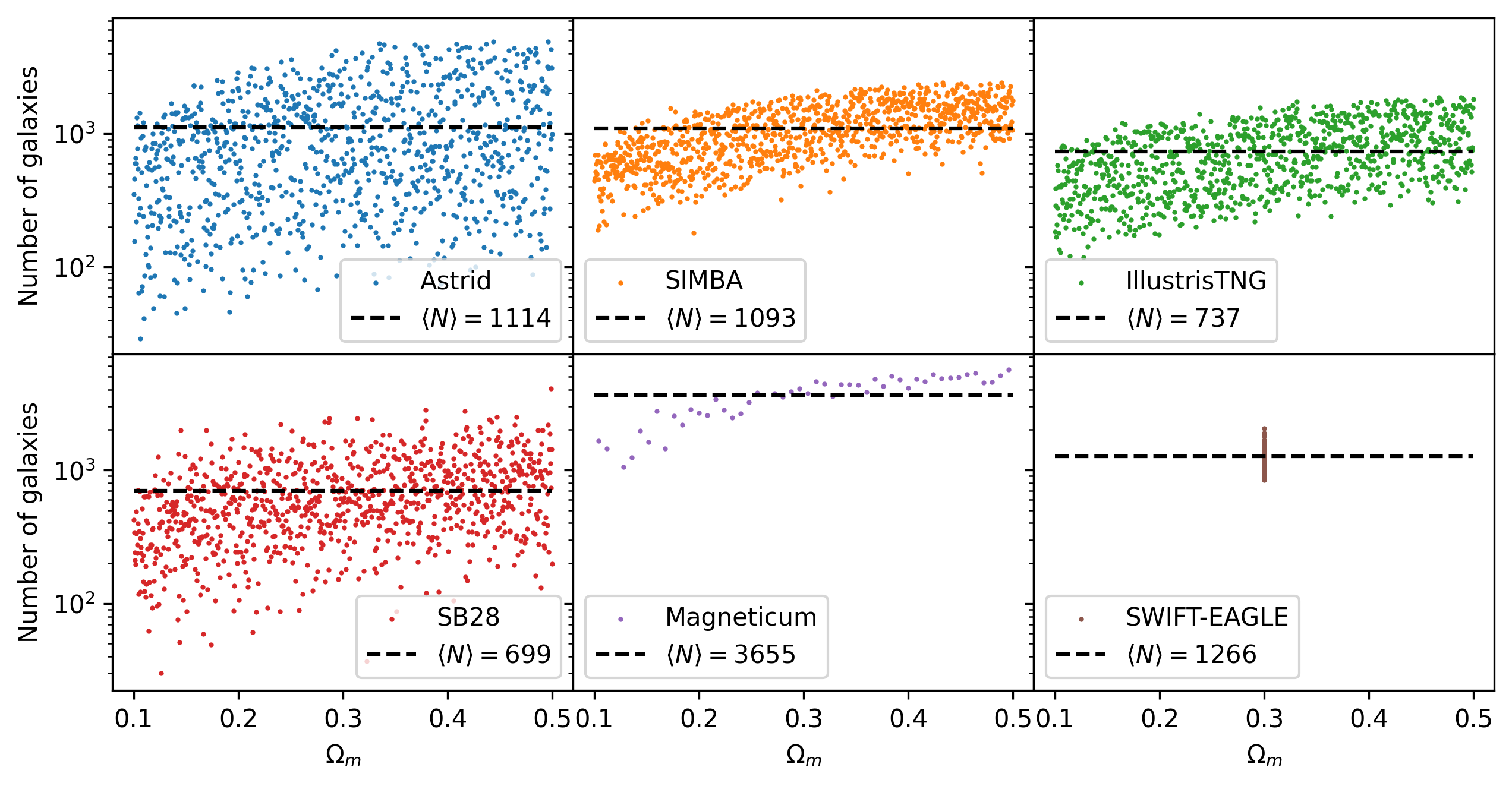}
 \caption{Comparison of the number of galaxies per LH catalog in CAMELS simulations for 
 Astrid (top left), SIMBA (top middle), IllustrisTNG (top right), SB28 (bottom left), 
 Magneticum (bottom middle), and SWIFT-EAGLE (bottom right). 
 The horizontal lines correspond to the mean number of galaxies per simulation.
 \label{fig:distros_CAMELS}
 }
\end{figure*}

All the graphs built in this work follow the prescription presented in Section 
\ref{sec:the_graph}. The simple visual inspection of Figure \ref{fig:graphs} indicates some
disparity among graphs from the different simulation suites. In this appendix, we explore some
other aspects of the graphs and their characteristics according to the different simulations.

In Figure \ref{fig:distros_CAMELS} we compare the number of galaxies in the LH 
catalogs for the different CAMELS simulations, considering a threshold in stellar mass as 
$M_{\star} = 1.95 \cdot 10^8 ~M_\odot/h$. 
In almost all the cases the mean number of galaxies is $\sim 1000$, being a bit lower
($\sim 700$) for IllustrisTNG and its variation SB28, and dramatically higher 
($\sim 3,600$) for Magneticum. 
Besides, we can see that Astrid includes catalogs with a huge range of galaxy number 
($N \in [30, 5,000]$), while the SIMBA and IllustrisTNG LH sets are much narrower (the same 
follows for SB28, with a higher dispersion of galaxy number, but not so broad as in 
Astrid). 
Finally, the range of the number of galaxies for Magneticum is $N \in [1000, 5500]$,
including catalogs with such a large number of galaxies that do not have equivalent simulations
in the SIMBA and IllustrisTNG data sets. 
As mentioned in Section \ref{sec:the_graph} the large number of galaxies in Magneticum is
related to the particular feedback model employed in those simulations.

The distances and the number of edges among the galaxies belonging to different catalogs were
also investigated, as well the percentage of single galaxies per catalog. 
As expected, the distances among galaxies cover a range 
$d \in [10^{- 2}, 21.65]~h^{-1}{\rm Mpc}$. 
All the catalogs have a similar shape in their spatial distributions, with small
differences on small scales.
Single galaxies (the ones which are not connected to any other, and therefore only
contribute to the propagation of their node information), on average, do not
correspond to more than $\sim 20 \%$ of the galaxies in the catalogs. This means that most of 
the information of the galaxies came from their connections (i.e. clustering properties). 
The number of edges per catalog is of order $\sim 10, 000$, indicating that most
galaxies have $\sim 10$ connections. 
Finally, the $r_{\rm link}$ found in all the models, for all different CAMELS sets in the
hyperparameter training optimization, was around $1.25~h^{-1}{\rm Mpc}$.

\section{Inferring sigma 8}
\label{sec:sigma8}

\begin{figure}
 \centering
 \includegraphics[scale=0.45]{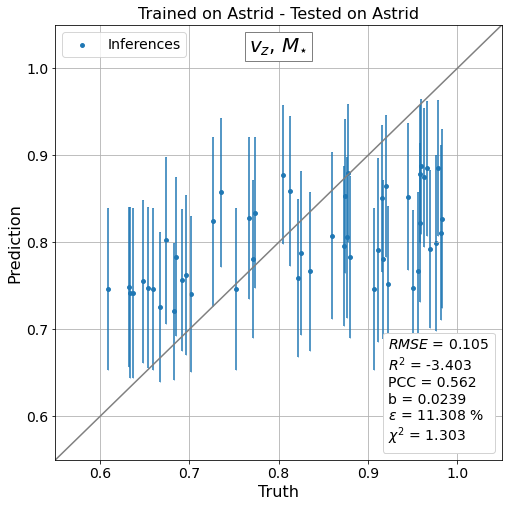}
 \includegraphics[scale=0.45]{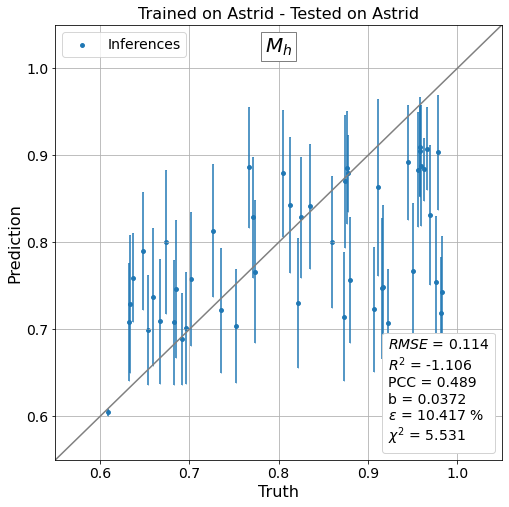}
 \caption{Likelihood-free inference of $\sigma_8$ using galaxy {\bf velocities on the $z$ 
 direction} and {\bf stellar mass} (on the left) and {\bf halo mass} (on the right) as node 
 attributes.
 We present the results for a model {\em trained} on {\bf Astrid} and {\em tested} on Astrid.}
 \label{fig:sigma_8}
\end{figure}

In this appendix, we present our efforts in trying to infer $\sigma_8$ using
galaxy catalogs as graphs to feed GNN models. We made a sequence of tests of properties to 
include as node information in our graphs and none of them resulted in a robust model. 
Here we present two main results guided by:
(1) \cite{pablo-galaxies-2022} while using galaxy velocities (in one direction) and 
including one more galaxy property, the stellar mass; and
(2) \cite{helen-halos-2022} when using the host halo mass as node information for the 
graphs. 
The results are shown in Figure \ref{fig:sigma_8}. 
In both models, we found poor performance: higher values for RMSE ($> 0.1$), negative values for 
$R^2$ ($- [3.4, 1.1]$) and low values for $PCC$ ($[0.49, 0.56]$). In the case of
the model which uses the halo mass, the $\chi^2$ value is higher too ($> 5.5$).
Furthermore, the predictions are around the fiducial/mean value, without covering the whole 
range of values and having higher error bars.

As already shown by \cite{pablo-galaxies-2022}, it is a challenge to infer this cosmological 
parameter using galaxy information, which may need more galaxy properties (stellar mass, 
galaxy radius, metallicity, and maximum circular velocity) to achieve better performance. 
Then, because of relying on galaxy properties that differ substantially among the different
simulations, it is hard to get a robust model. 
That is why our inference while using only galaxy velocity and stellar mass, is 
worse than these authors' results. 
On the other hand, because we are using all galaxies (centrals and satellites), our 
results are not directly comparable to the ones presented in \cite{helen-halos-2022}, where 
only halo catalogs (without subhalos) are employed. 

Therefore, we conclude, in agreement with \cite{pablo-galaxies-2022} and \cite{one_gal-2022} that to 
constrain $\sigma_8$ precisely, we need larger volumes, as no ML technique was able to 
infer their value using only galaxy information.
Besides, getting the correct value of this parameter can be challenging also for the standard 
approaches due to the small size of the boxes in the CAMELS suite.
One possible solution can be found in \cite{lucia2022}, where the authors obtained good constraints 
to predict $\sigma_8$ using machine learning methods to deal with the usual summary statistics, for
larger boxes ($100 ~h^{- 1}$Mpc).
Another possible way to solve the puzzle related to $\sigma_8$ predictions should
train a GNN on galaxy catalogs at higher redshifts and look for their impact on galaxy
populations. This can be mostly related to the response of $\sigma_8$ in the abundance
of more massive structures due to hierarchical structure formation, which does not happen at
$z = 0$, where small galaxy populations dominate \citep{Yueying-prep}. This will be addressed 
in future work.

\section{SIMBA and IllustrisTNG results}
\label{sec:SIMBA_IllustrisTNG}

\begin{figure}
 \centering
 \includegraphics[scale=0.4]{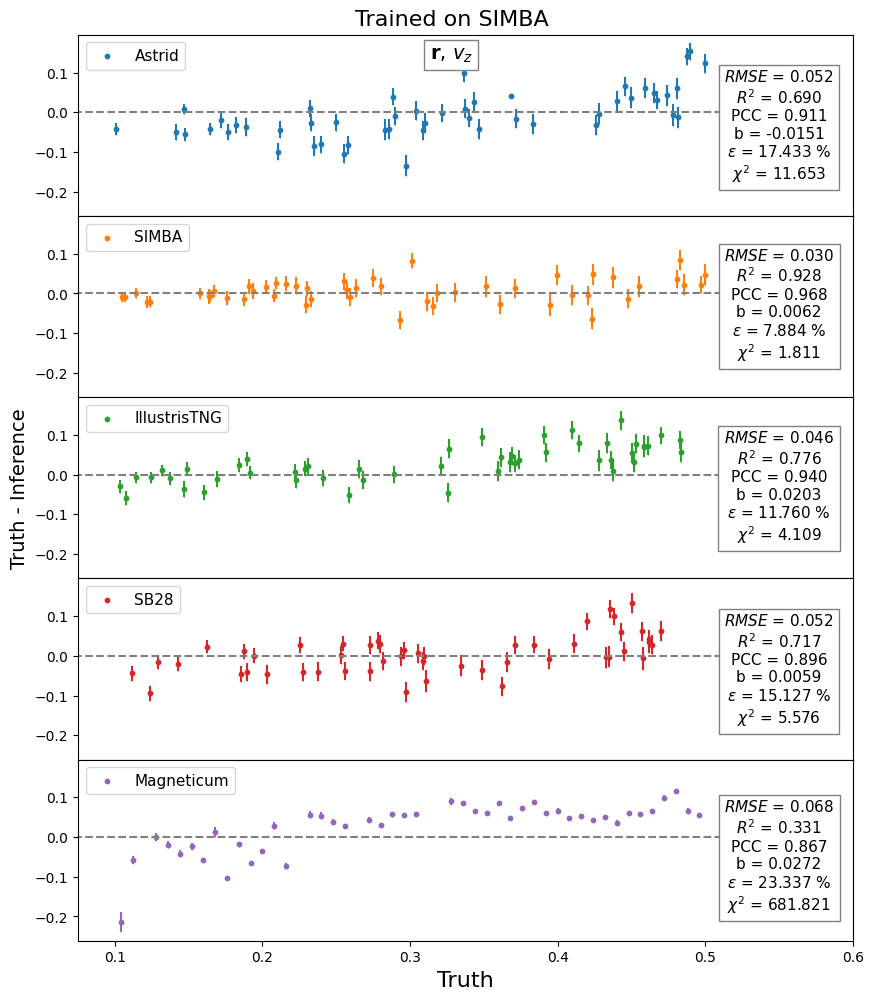} 
 \includegraphics[scale=0.4]{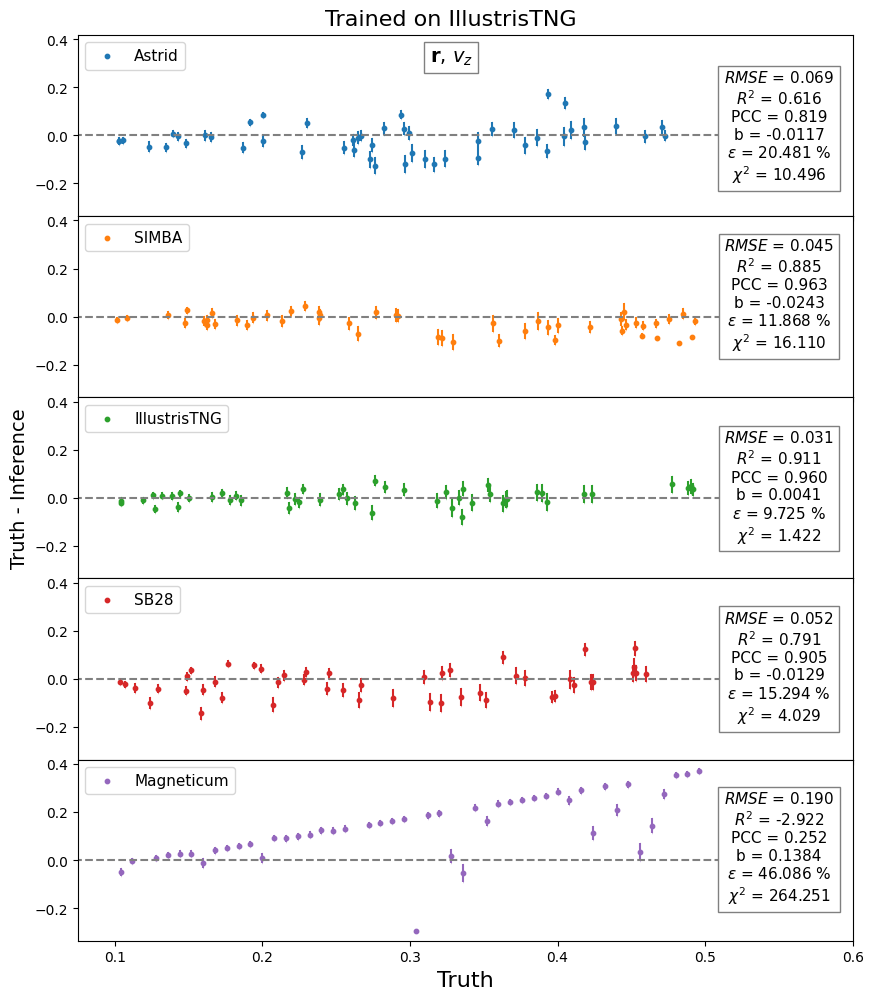} 
 \caption{Likelihood-free inference of $\Omega_{\rm m}$ using galaxy {\bf positions} and 
 {\bf velocities in the $z$ direction}. We present the results for {\bf LH set} tests of a 
 model {\em trained} on {\bf SIMBA} (on the left) and {\bf IllustrisTNG} (on the right) 
 and {\em tested} on Astrid, SIMBA, IllustrisTNG, SB28, and Magneticum respectively from 
 the top to the bottom.}
 \label{fig:LH_SIMBA_IllustrisTNG}
\end{figure}

\begin{figure}
 \centering
 \includegraphics[scale=0.34]{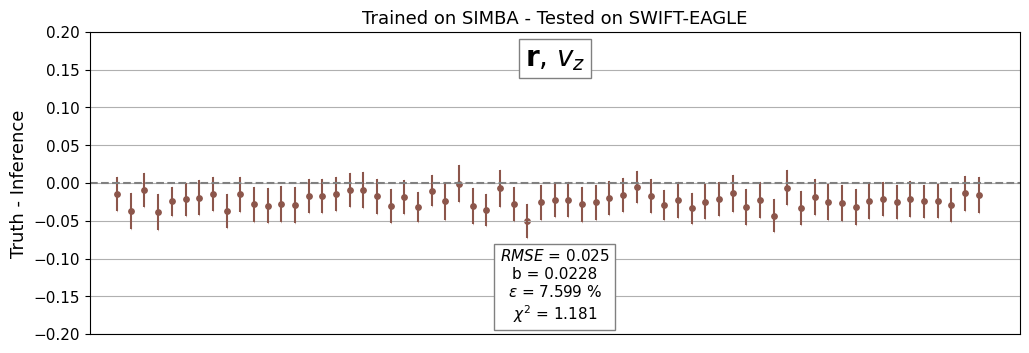}
 \includegraphics[scale=0.34]{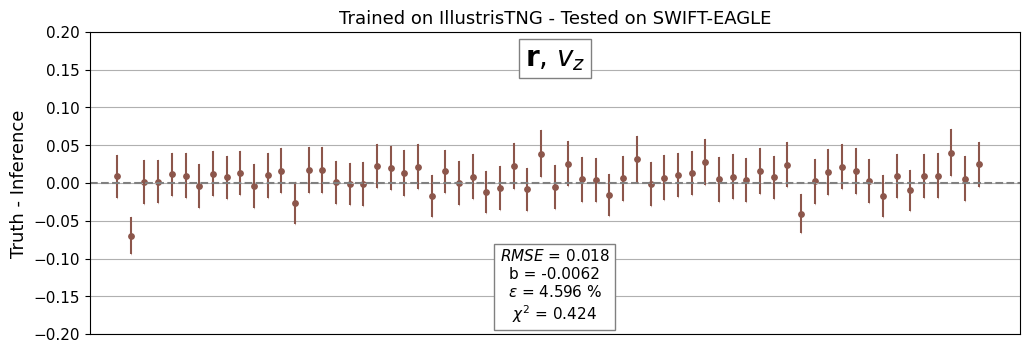}
 \caption{Likelihood-free inference of $\Omega_{\rm m}$ using galaxy {\bf positions} and 
 {\bf velocities in the $z$ direction}. We present the results for a model {\em trained} on
 {\bf SIMBA} (on the left) and {\bf IllustrisTNG} (on the right) and {\em tested} on 
 SWIFT-EAGLE.}
 \label{fig:SWIFT_SIMBA_IllustrusTNG}
\end{figure}

\begin{figure}
 \centering
 \includegraphics[scale=0.34]{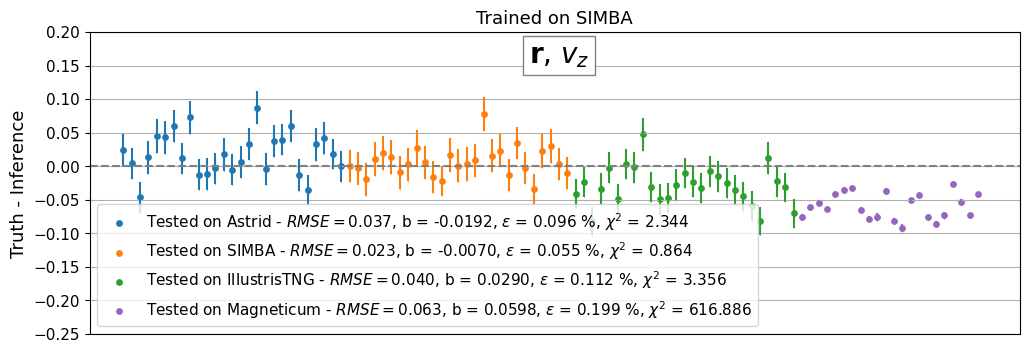}
 \includegraphics[scale=0.34]{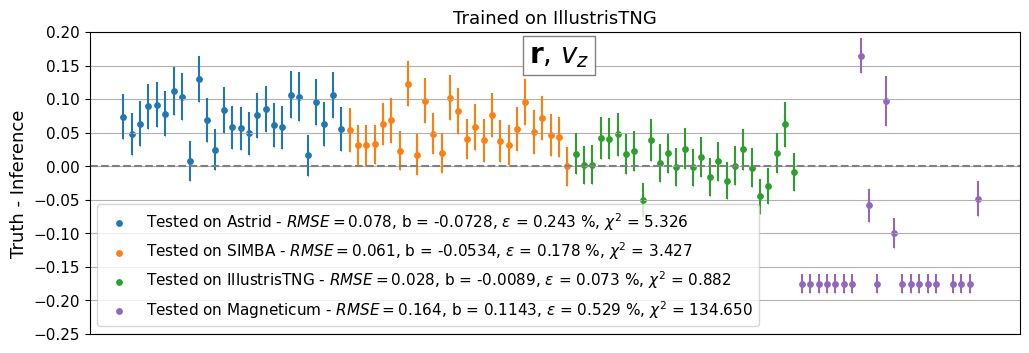}
 \caption{Likelihood-free inference of $\Omega_{\rm m}$ using galaxy {\bf positions} and 
 {\bf velocities in the $z$ direction}. We present the results for {\bf CV set} tests of a 
 model {\em trained} on {\bf IllustrisTNG} and {\em tested} on Astrid, SIMBA, 
 IllustrisTNG, SB28, and Magneticum.}
 \label{fig:CV_SIMBA_IllustrusTNG}
\end{figure}

The present appendix follows the results of Section \ref{sec:Astrid-best_model}, for 
models trained using SIMBA and IllustrisTNG data sets. We stress that the GNN architecture 
follows the same structure as the one used in the best model (but with a different set of 
hyperparameters, found using \textsc{Optuna}). 

All the results are presented in Figures \ref{fig:LH_SIMBA_IllustrisTNG}, 
\ref{fig:SWIFT_SIMBA_IllustrusTNG} and \ref{fig:CV_SIMBA_IllustrusTNG}, where we plot the 
values for \{truth - inference\} in the $y$-axis, while the $x$-axis shows either the truth values 
of $\Omega_{\rm m}$ or an arbitrary order of the predictions by simulation suite. 
The metrics for the models trained on SIMBA/IllustrisTNG and tested on themselves 
are very good (even compared to the best model): $RMSE = [0.030, 0.031]$, 
$R^2 = [0.911, 0.928]$, $PCC = [0.960, 0.968]$, $b = [0.0041, 0.0062]$, 
$\epsilon = [7.9, 9.7] \%$, and $\chi^2 = [1.422, 1.811]$. However, all the tests on the other 
simulations are worse: 
$RMSE \in [0.018, 0.190]$, $R^2 \in [- 2.922, 0.885]$, $PCC \in [0.252, 0.963]$, 
$b \in [0.0059, 0.1384]$, $\epsilon \in [4.6, 46.1] \%$, and 
$\chi^2 \in [0.424, 681.821]$.
The worst predictions show up when the networks are tested on Magneticum (both, for the model 
trained on SIMBA and IllustrisTNG, but being worse for the latter). The tests on SWIFT-EAGLE 
and in the CV sets show that the scores are, in most cases, a bit worse compared to the best 
model (when we train the model using Astrid).

Our results suggest that the very poor predictions for Magneticum are due to the fact that
the models trained on SIMBA and IllustrisTNG have never seen catalogs 
with such a high number of galaxies, which is the case for Magneticum catalogs (see 
Appendix \ref{sec:graph_details}, specially Figure \ref{fig:distros_CAMELS}, which shows 
that Astrid covers a large range of number of galaxies when compared to SIMBA and 
IllustrisTNG). We have tested to increase the stellar mass cut in Magneticum catalogs and 
have obtained better predictions (comparable to the same models tested on the other catalogs 
apart themselves) while using the models trained on SIMBA/IllustrisTNG. 
This shows that reducing the number of galaxies in Magneticum catalogs improves their 
inferences significantly. 
Therefore, although the number of galaxies is not the most important property in the 
analysis, we can clearly see their effect on the model predictions while taking a look at
these results.

Finally, in contrast to the robust model that was trained using Astrid, the inferences
from the models trained using SIMBA and IllustrisTNG are, unfortunately, not robust across
different simulations.

\section{Super-sample covariance analysis}
\label{sec:TNG300}

We start noticing that our $25~h^{-1}{\rm Mpc}$ boxes have a mean overdensity, 
$\langle \rho/\bar{\rho}\rangle=1$. In the real Universe, $(25~h^{-1}{\rm Mpc})^3$ patches 
will not satisfy that equality, and values larger or smaller will appear due to the presence 
of power on modes larger than the size of that region. Those modes are expected to affect both
the clustering of galaxies and their internal properties. Here we investigate whether such 
effects will affect our models. To test this, we made use of the IllustrisTNG300-1 simulation, 
which covers a periodic volume of $(205~h^{-1}{\rm Mpc})^3$ at a slightly higher resolution 
than the CAMELS simulations. 

\begin{figure}
 \centering
 \includegraphics[scale=0.34]{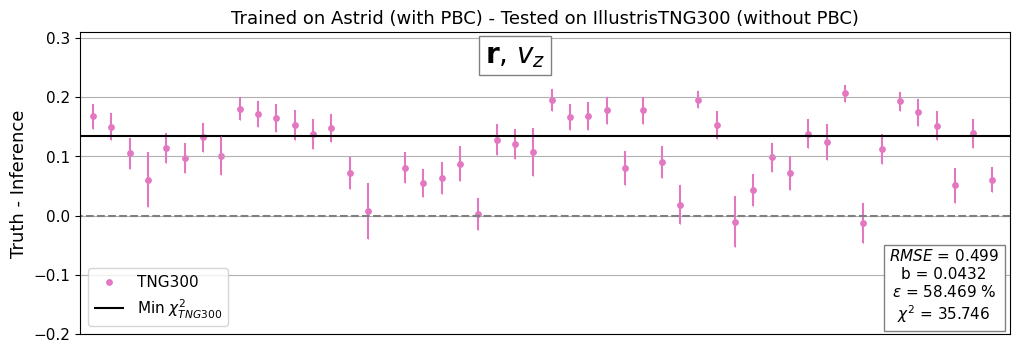}
 \includegraphics[scale=0.34]{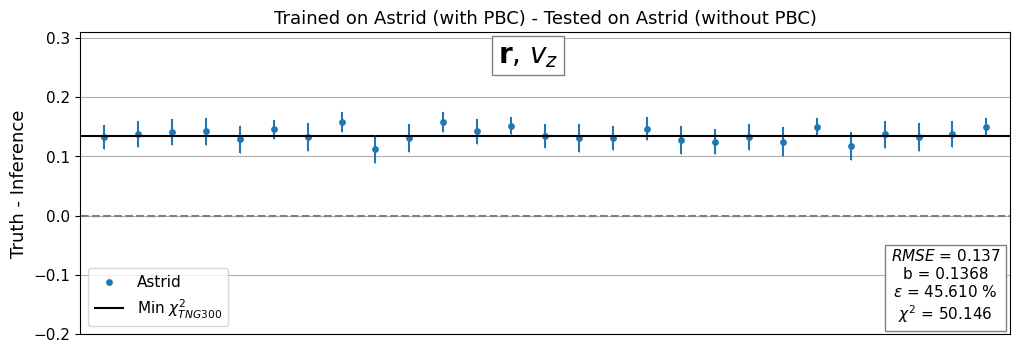}
 \caption{Likelihood-free inference of $\Omega_{\rm m}$ using galaxy {\bf positions} and 
 {\bf velocities in the $z$ direction}. We present the results for a model {\em trained} on 
 {\bf Astrid} and {\em tested} on: (1) $50$ random $(25~h^{-1}{\rm Mpc})^3$ sub-volumes of the
 {\bf IllustrisTNG300} simulation (on the left) and (2) Astrid (on the right). In both cases 
 the model was trained considering the periodic boundary conditions (PBC) and tested without 
 this consideration.}
 \label{fig:TNG300}
\end{figure}

We have selected $50$ random $(25~h^{-1}{\rm Mpc})^3$ sub-volumes within the IllustrisTNG300 
box, taking the galaxies in those sub-volumes and constructed graphs to input into our model. 
It is important to note that we have turned off the periodic boundary conditions (PBC) when
constructing the graphs, due to the fact that the distribution of galaxies is not periodic 
within the sub-volumes. The results of testing our model with these galaxy catalogs are shown 
in Figure \ref{fig:TNG300}. We can see that the inferences for the IllustrisTNG300 catalogs 
have a positive bias of $b = 0.0432$ and the different estimations fluctuate around an offset 
that we indicate as Min $\chi^2_{\mathrm{TNG300}}$. This value represents the $\chi^2$
minimization considering the IllustrisTNG300 inferences.

In order to test if this offset can be an effect of turning off the PBC we have tested the 
model on Astrid galaxy catalogs whose graphs have been constructed neglecting PBC. The results 
are presented in the right panel of Figure \ref{fig:TNG300}. We can see that we find almost the 
same offset for these new predictions.

\begin{figure}
 \centering
 \includegraphics[scale=0.45]{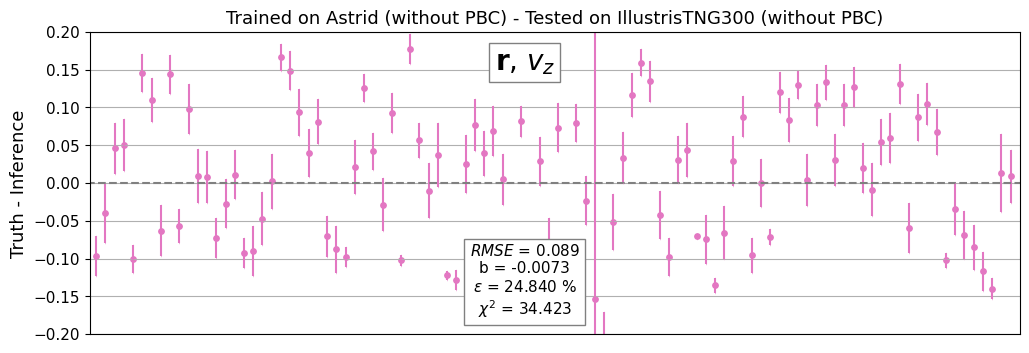}
 \caption{Likelihood-free inference of $\Omega_{\rm m}$ using galaxy {\bf positions} and 
 {\bf velocities in the $z$ direction}. We present the results for a model {\em trained} on 
 {\bf Astrid} and {\em tested} on $100$ random $(25~h^{-1}{\rm Mpc})^3$ sub-volumes within
 {\bf IllustrisTNG300}. This specific model was trained without the periodic boundary 
 conditions (PBC) and tested without this too.}
 \label{fig:2TNG300}
\end{figure}

Given the large effect that the PBC have on our results, we have retrained the GNN model on Astrid 
galaxy catalogs whose graphs are constructed without using PBC. We then test that model on 
galaxy catalogs from $100$ random sub-volumes of the IllustrisTNG300 simulation. The results 
are presented in Figure \ref{fig:2TNG300}. We can see that the inferences do not exhibit good 
scores: $RMSE = 0.089$, $b = - 0.0073$, $\epsilon = 24.8 \%$, and $\chi^2 = 34$. Even so, all 
the predictions fluctuate around the true values, indicating that we may have outliers.
After removing predictions related to $\chi^2 > 14.0$ ($35$ points) we achieve
better results that follows for: $RMSE = 0.059$, $b = - 0.0118$, $\epsilon = 16.6 \%$, and 
$\chi^2 = 4,0$.

From these results, we conclude that our method is not severely affected by super-sample 
covariance in the majority of the cases, although it does not work in all scenarios. We note 
that the fraction of outliers (i.e. cases where the model performs badly) is much higher in 
this test case than in, e.g., SB28 simulations. This indicates that further work is needed to 
either account for super-sample covariance effects or to identify the range of validity of our 
models. We leave this task for future work.

\bibliography{sample631}{}
\bibliographystyle{aasjournal}



\end{document}